\documentclass[12pt]{article}
\usepackage[english]{babel}
\usepackage{geometry}
\usepackage{xcolor}
\usepackage{fancyhdr}
\usepackage[labelfont=bf]{caption}
\usepackage[T1]{fontenc}
\usepackage{lipsum}
\usepackage{amsmath}
\usepackage[colorinlistoftodos]{todonotes}
\usepackage{float}
\usepackage{subcaption}
\usepackage[stable]{footmisc}
\usepackage{hyperref}
\usepackage{hyperref}
\usepackage{array}
\newcolumntype{C}[1]{>{\centering\arraybackslash}p{#1}}
\usepackage{longtable}
\usepackage{tabularx}
\usepackage[sorting=none]{biblatex}
\usepackage{xcolor}
\date{}
\title{\textbf{Layer 2 Blockchain Scaling: a Survey}}

\author{
		Cosimo Sguanci\\
		\small\color{blue}\href{mailto:cosimo.sguanci@mail.polimi.it}
			{\nolinkurl{cosimo.sguanci@mail.polimi.it}}
	\and 
		Roberto Spatafora\\
		\small\color{blue}\href{mailto:roberto.spatafora@mail.polimi.it}
			{\nolinkurl{roberto.spatafora@mail.polimi.it}}
	\and 
		Andrea Mario Vergani\\
		\small\color{blue}\href{mailto:andreamario.vergani@mail.polimi.it}
			{\nolinkurl{andreamario.vergani@mail.polimi.it}}
}

\date{June 15, 2021\\\small Version 1.0}

\begin{document}
\maketitle


\thispagestyle{empty}
\begin{abstract}
  \noindent Blockchain technology is affected by massive limitations in scalability with consequent repercussions on performance. This discussion aims at analyzing the state of the art of current available Layer II solutions to overcome these limitations, both focusing on theoretical and practical aspects and highlighting the main differences among the examined frameworks. The structure of the work is based on three major sections. In particular, the first one is an introductory part about the technology, the scalability issue and Layer II as a solution. The second section represents the core of the discussion and consists of three different subsections, each with a detailed examination of the respective solution (Lightning Network, Plasma, Rollups); the analysis of each solution is based on how it affects five key aspects of blockchain technology and Layer II: scalability, security, decentralization, privacy, fees and micropayments (the last two are analyzed together given their high correlation). Finally, the third section includes a tabular summary, followed by a detailed description of a use-case specifically thought for a practical evaluation of the presented frameworks. The results of the work met expectations: all solutions effectively contribute to increasing scalability. A crucial clarification is that none of the three dominates the others in all possible fields of application, and the consequences in adopting each, are different. Therefore, the choice depends on the application context, and a trade-off must be found between the aspects previously mentioned.
 
\end{abstract}

\section{Blockchain Introduction}
In its most basic meaning, Blockchain technology consists of a decentralized public ledger capable of keeping a record of immutable (and linked) information. However, the original goal of this technology, as stated in the original Bitcoin white paper written by Satoshi Nakamoto, is to provide a trustless electronic payment system that does not rely on any third-party authority to manage disputes between actors. The trust, which is needed in traditional cash systems, can be replaced by cryptographic proofs. 
\par Such a system can be implemented as a {\textit{peer-to-peer}} network of nodes, in which a certain number of transactions are periodically bundled into blocks. Blocks are hashed to generate timestamps and every block includes the hash of the previous block. This mechanism forms a chain of timestamped information that, in this context, is represented by transactions. All the generated blocks are propagated across the network and locally stored by each node; in this way every node can act as a {\textit{validator}} of transactions, and the so-called {\textit{double spending}} problem can be solved.
\par To allow {\textit{peer-to-peer}} direct payments, transactions must be computationally infeasible to reverse: this is allowed by the {\textit{proof-of-work}} mechanism, in which every node interested in generating new blocks (also called {\textit{miners}}) search for a value that, when hashed using a particular hash algorithm, has certain properties. For instance, as in the Bitcoin protocol, this could involve the fact that the resulting hash must have a certain number of leading zeros. {\textit{Miners}} are then incentivized to sustain the network by collecting the fees that users spend to execute transactions, thus improving the global {\textit{hashrate}} of the Blockchain.
\par Thanks to this paradigm, combined with the fact that nodes consider the longest chain to be the valid state, once a block has been generated and the computation has been performed, changing the block would require redoing the {\textit{proof-of-work}} for both the specific block and all subsequent ones.
\par This system has proven to be secure as soon as honest nodes hold the majority of the network computational power: in this case, honest participants will generate the longest chain, which will be accepted as the true state of the Blockchain, not allowing attackers to create malicious blocks.
\par It is common to identify three different generations of Blockchains:

\begin{itemize}
\item First generation: Blockchains that allow decentralized monetary system and ledger of transactions. An example is the \textit{Bitcoin} Blockchain.
\item Second generation: Blockchains with support for running {\textit{smart contracts}} (e.g. \textit{Ethereum}). These Blockchains are decentralized platforms which can be used to run programs in a decentralized fashion, to achieve complex functionalities without the need for a central authority (such as an Application Server).
\item Third generation: Blockchains that are exploring different {\textit{consensus algorithms}} other than {\textit{proof-of-work}} (one example is the {\textit{proof-of-stake}} mechanism). They generally also aim at substantially improving the scalability of pre-existing Blockchains. Examples of this type of Blockchain are \textit{Cardano}, \textit{Polkadot}.
\end{itemize}

\section{Blockchain Scalability issue}
A critical aspect of the blockchain concerns its scalability. At the moment of writing, scalability is considered the bottleneck of the blockchain infrastructure. The aforementioned could potentially compete with the largest electronic payment circuits. However, it is limited by being able to handle few transactions per second (TPS). To clarify the problem, it is sufficient to report the two best-known blockchains as an example: Bitcoin processes 4.6 TPS and Ethereum processes around 14.3 TPS (slightly variable values), while one of the largest electronic payment circuits, Visa, processes around 1,736 TPS (and has been able to reach peaks of 47,000 TPS). \\ Currently, the Bitcoin blockchain generates a new block every about 10 minutes (Time Block generation, TB) and, the block size (B) in the chain is 1MB (1,048,576 Bytes). The average transaction size is 380 Bytes. Therefore, the number of transactions that fit into a Bitcoin block (TPB) is: $$TPB = \frac{Block \;Size}{Average\;Transaction\;Size} = \frac{1,048,576\;Bytes}{380\;Bytes}\approx 2,759\;transactions$$As a consequence of the TB and TPB, the number of transactions per second is: $$TPS = \frac{TPB}{TB}\approx\frac{2,759\;transactions}{600\;seconds}\approx4.6\;transactions\;per\;second$$ 
\par Contrary to what one might think, the scalability problem cannot be solved by simply modeling its parameters: B and TB. As a matter of fact, considering the parameters modeling, it is necessary to make an important clarification: when generating a new block in the blockchain, a crucial factor to be taken into account is the relay time (TR) needed to broadcast the new block to every node on the network. Therefore, this fact imposes a lower limit on TB, below which it is impossible to go. This TR threshold allows to keep all the nodes in the network constantly updated. Additionally, another problem related to the TR arises when expanding the block size (B). Consequently, an increased quantity of information has to be broadcasted to the network.\\ Again, a practical example from Bitcoin blockchain is considered in the discussion for simplicity of presentation: in January 2021,  around 10,000 nodes were estimated in the Bitcoin network. The average time to propagate a block to 99\% of the network is approximately 14 seconds. Thus, TB cannot fall below the 14 seconds threshold. Otherwise, a new block would be generated before an old block would be received by most of the nodes in the network. The problem related to the size of the block becomes evident when increasing it: 14 seconds as Time Relay would no longer be enough. 
\par In 2017, Segregated Witness’s (SegWit) soft fork helped to improve block size (scaled up theoretically to 4MB, practically around 2MB) without changes to the core code. Nevertheless, it still does not improve TPS in a scalable manner.


\section{Layer II solutions}
As explained in the previous section, blockchain's main problem is scalability; this issue derives from the fact that the technology is driven by a decentralized idea as its core, which makes it difficult to scale since transactions have to be broadcasted to the whole network. 
\par All solutions to the presented issue have to deal with the so-called {\textit{scalability trilemma}}: improvements on blockchain's scalability have a negative effect either on security or decentralization, or both. Since having a decentralized secure network is one of the pillars of blockchain technology, a right tradeoff needs to be found in order to scale. 
\par The most common and used approach to achieve a scalable blockchain is generally known as ``Layer 2'': the basic idea is that of building a framework which handles transactions {\textit{off-chain}} (not on the main chain and, in a certain sense, independently of it), thus reducing the load on the blockchain itself and achieving higher transaction speed. Of course, as hinted above, moving in the direction of scalability, and in particular registering transactions {\textit{off-chain}}, leads to problems in terms of security and decentralization, which need to be addressed by specific countermeasures. 
\par A Layer 2 solution is a secondary protocol built on-top of an existing blockchain; the idea behind the framework can be of different natures, but the key-concept is that of hosting transactions and reporting only a ``summary'' of them on the main chain. To better understand the situation, it is necessary to distinguish various kinds of high-level Layer 2 (L2) solutions:
\begin{itemize}
\item Channels L2 basically create direct or indirect {\textit{off-chain}} communication channels between nodes; transactions between ``connected'' nodes are managed on Layer 2, reporting on the main chain only two of them (the one ``opening'' the channel and the one which ``closes'' it). For example, Lightning Network for Bitcoin and Raiden Network for Ethereum are based on state channel L2.
\item Sidechains L2 are based on ``children'' blockchains anchored on the main chain and running in parallel to it; the idea is, in a sense, quite similar to channels, but the consistent difference is that, in sidechains, {\textit{off-chain}} transactions run on blockchains (while communication channels are not based on a blockchain). Of course, the advantage of moving transactions to another blockchain, which in principle could suffer from the same scalability issue as the main chain, is the following: sidechains involve less nodes and typically ``weight'' in a different way the {\textit{trilemma}} between scalability, security and decentralization (in general, they tend to be less decentralized and faster). In the dedicated section, there will be a distinction between standard sidechains and Plasma, a sidechain architecture that offers more guarantees in terms of decentralization and security. Again, transactions handled by a sidechain are reported {\textit{on-chain}} with only an ``opening'' and a ``closing'' one.
\item Rollups L2 lie on the general concept of only {\textit{executing}} a transaction {\textit{off-chain}}, but always reporting data about it on the main chain; in practice, while channels and sidechains need to report the ``summary'' of a set of transactions with just two of them on the blockchain, Rollups broadcast a smaller amount of data (with respect to the usual size of {\textit{on-chain}} transactions) for every off-chain state update.
\end{itemize}
It is easy to understand that the main concept of all layer 2 solutions is that of lightening the blockchain, in order to help in scaling up. The consequence of this idea is not only higher transaction speed, but also (in general) lower fees (which is a direct consequence of increased TPS): this means that L2 solutions are appropriate places for micropayments to be performed. Other advantages of L2 are the facts that no modification on the main chain is needed and that {\textit{off-chain}} management of transactions is, in a sense, independent of ``Layer 1''; in reality, dependence is essential in order to register {\textit{on-chain}} a ``summary'' of transactions, but, apart from this fact, the blockchain is not aware of what happens on Layer 2. 
\par An honourable mention for what regards blockchain scaling solutions is given to sharding. It is a technique based on partitioning the main chain into subsets of nodes, each responsible for a portion of the whole network: every node processes information belonging only to its shard. This solution certainly goes in the direction of improving scalability: ``dividing'' the load of the chain among different partitions leads to something similar to separate blockchains, characterized by higher transaction speed (since they are lighter); of course, the main chain is not really divided into smaller chains, because shards are still able to share information; however, sharding clearly goes in the direction of lower security and decentralization, thus enforcing once again the {\textit{scalability trilemma}}. Sharding can not be properly defined as a Layer 2 solution, since no additional {\textit{off-chain}} framework is effectively added to the main chain; for this reason, this technique is generally referred to as ``Layer 1 scaling solution'', to mean exactly the fact that all transactions are managed on the blockchain itself. 
\par The last remark about L2 frameworks is that different protocols can be combined, on-top of the same main chain, in order to improve scalability as much as possible. This is feasible thanks to the fact that the blockchain is totally unaltered and not affected by the use of Layer 2 solutions.

\section{Lightning Network (and references to Raiden Network)\footnote{Raiden Network is an in-development technology, with less than 65 open channels at the moment of writing. For this reason, this section is mainly focused on the description of Lightning Network. For what regards Raiden Network, descriptions and hints are provided only for the sake of completeness and/or in case of difference with respect to Lightning Network functioning; instead, when nothing else is mentioned, the reader can deduce a very similar behaviour for both Lightning Network and Raiden Network. Since Lightning Network is going to be described also in relation to its actual topology, the reasonable assumption (if not differently specified) is that Raiden Network is likely to evolve in a similar manner in its near future; however, nothing more than assumptions can be carried on at the moment of writing under this point of view, given Raiden Network immaturity.}}\label{sec:LN}
Lightning Network (LN) and Raiden Network (RN) are channel networks that operate on top of a blockchain. They have the main objective of enabling fast \textit{peer-to-peer} transactions. LN and RN functioning is based on state channels: the main idea behind this type of solution is the creation of an \textit{off-chain} communication channel between nodes. The L2 solution is in charge of managing transactions between directly or indirectly connected nodes, therefore reducing the main chain's workload, which only has to track the channel's opening and closing transactions. Thus, it allows transactions among nodes in the network which update the tentative distribution of the channel's funds without broadcasting those to the blockchain. LN was introduced as a Layer 2 solution for the Bitcoin blockchain, but it has been integrated also by other blockchains (\textit{e.g.}, Litecoin)\footnote{From this moment on, Lightning Network description is focused on its use as Bitcoin blockchain Layer 2 solution.}; on the other side, RN is Lightning Network's counterpart allowing to perform \textit{off-chain} payments for Ethereum.
\par Lightning Network has a separate structure from the main chain, with which it communicates only special opening and closing channel transactions. The opening transaction creates a ``channel'' between two nodes in the network. Once there is a channel, all the transactions between the two parties are stored in a ``private ledger''. Only the two counterparties can access the ledger, and neither party can cheat on the other one. The impossibility of cheating is a significant aspect that will be better analyzed in the discussion. 
\par To clearly understand the use of the channel, let's imagine two users, Alice and Bob, who opened one. Each of them put 3 BTC in a ``contract'' with the counterparty. After ten transactions, a possible scenario might be: Alice has 5 BTC on her side, and Bob has 1 BTC on his side.
\begin{figure}[H]
\begin{subfigure}{.5\textwidth}
\raggedright
\includegraphics[width=7.5cm]{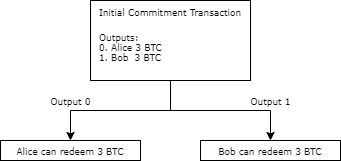}
\caption{Initial scenario}
\end{subfigure}\hfill
\begin{subfigure}{.5\textwidth}
\raggedleft
\includegraphics[width=7.5cm]{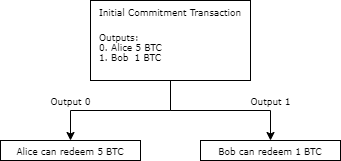}
\caption{Scenario after 10 transactions}
\end{subfigure}\hfill
\end{figure}
\noindent At any time, either party can publish the current state on the blockchain. At that point, the balances on each side of the channel are allocated to their respective parties \textit{on-chain}. In Lightning Network, this implies the closure of the channel; in Raiden Network, instead, it is also possible for a user to ``withdraw tokens from a channel without closing it''\footnote{Quotation taken from ``Withdraw tokens from a channel'' section of \url{https://raiden-network-specification.readthedocs.io/en/latest/smart_contracts.html}} (having his/her counterparty's signature as well).
\par Transactions in the channel are not subject to fees except for the channel opening and closing ones. As a consequence, a substantial advance of LN is that it allows micropayments (see \hyperref[sec:FeesAndMicropaymentsLN]{section~\ref{sec:FeesAndMicropaymentsLN}: \textit{Fees and micropayments}}). In particular, since transactions in the channel have no fees (without considering payment routing, described in the following sections), users can be encouraged to send even small amounts of money. Thus, it is now possible to transact the smallest unit currently available, 0.00000001 BTC (one \textit{satoshi}).
\par Another relevant feature of LN is the privacy it gives to users. Except for the first and last transactions, nobody outside the channel can see what happens inside it (see \hyperref[sec:PrivacyLN]{section~\ref{sec:PrivacyLN}: \textit{Privacy}}). 

\subsubsection*{Multisignature addresses \& HTLC}
A multisignature (multisig) address is one that multiple private keys can spend from. When creating the channel, the number of private keys that can spend funds and how many of those keys are required to sign a transaction are specified. For instance, a 2-of-4 scheme would indicate that, on four possible keys, any two of them are required to authorize a transaction (signing it). It is substantial to point out that, when the multisig needs more than a single key to authorize a transaction, single users cannot move funds without other users agreeing. \\A simple example: Alice (A) and Bob (B) lock up 5 BTC each into a 2-of-2 scheme. This scheme is composed of only two private keys capable of signing, and both are needed to move money. Supposing that A ends up with 9 BTC and Bob with 1 BTC, this scenario might lead Bob not to cooperate, locking his and Alice's funds in the multisig address (note that Alice needs Bob agreeing to unlock her funds). A mechanism that prevents the problem that might occur when one of the parties decides not to cooperate goes under the name of Hash TimeLock Contract (HTLC). In the \textit{Security} section that follows shortly, the process by which Lightning Network prevents cheating is explained in detail.

\subsubsection*{Routing payments}
A substantial part of the usefulness of LN is due to the connection between channels. The existence of different paths allows payments between users even if those are not directly connected. For instance, if Alice (A) opens a channel with Bob (B), and Bob already has one with Carol (C), Bob can act as intermediary by routing payments between A and C.
\begin{figure}[H]
\centering
\includegraphics[width=0.6\textwidth]{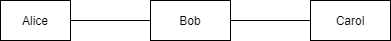}
\end{figure}
Intermediaries receive coins in the channel with the sender. Then, they effectively spend their funds in the route toward the receiver. A significant clarification is that the overall amount of money in a channel remains the same. The following representation extends the example above, with Bob routing a payment between Alice and Carol. The reported scenario shows a channel between Alice with a 5 BTC balance and Bob with a 2 BTC balance; the second channel, between Bob with 3 BTC and Carol with 1 BTC, is reported too. Supposing Alice wants to send 2 BTC to Carol, she should send 2 BTC to Bob; then, Bob, from the other channel, should send 2 BTC to Carol. The final scenario would be: Alice 3 BTC, Bob 4 BTC (Alice side), Bob 1 BTC (Carol side), Carol 3 BTC. 
\begin{figure}[H]
\centering
\includegraphics[width=0.7\textwidth]{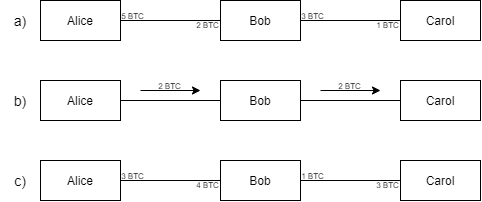}
\caption*{\textbf{Figure:} a) Initial scenario; b) Balance flows; c) Final scenario}
\end{figure}
The importance of routing payments in state channels Layer 2 solutions is exemplified by the fact that in Raiden Network, for instance, every payment, under the implementation point of view, is seen as a multi-hop transaction with N intermediaries; if the channel supporting it is direct, then N=0.
\par A criticism of routing payments is that it is impossible to spend more than the fund locked in a channel. Therefore, a considerable problem might occur when, considering the same example as before, Alice wants to send all her funds (5 BTC) to Carol. This transaction cannot be performed with Bob as an intermediary because Bob has just 3 BTC (Carol's side), which correspond to the maximum amount of money Alice can send to Carol via Bob. 
\begin{figure}[H]
\centering
\includegraphics[width=0.7\textwidth]{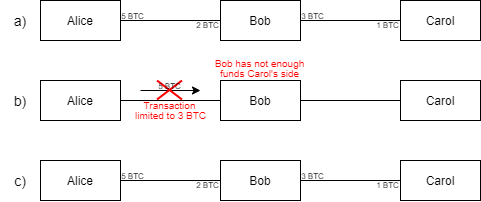}
\caption*{\textbf{Figure:} a) Initial scenario; b) Transaction denied; c) Final scenario as the initial state.}
\end{figure}
A relevant aspect to note is that, even if it is not mandatory, intermediaries usually ask for fees to route transactions. In particular, since they effectively spend their funds in involved channels, they ask senders to pay a fee (a small fee compared to that of the main chain) for having their payments routed to the recipient. This aspect might imply relevant consequences for the possibility of micropayments in a not fully decentralized solution, as it is further detailed in \textit{Decentralization} and \textit{Fees and micropayments} sections.


\subsection{Scalability}
\label{sec:ScalabilityLN}
Layer 2 solutions aim at solving blockchain's inherent scalability problem. Lightning Network achieves this objective for Bitcoin's blockchain by increasing from almost 5 TPS to (theoretically) billions of transactions per day, with an estimated average of at least 11,000 TPS. It enables transactions to be confirmed instantly, securely 
(see \hyperref[sec:SecurityLN]{section~\ref{sec:SecurityLN}: Security}),
maintaining a decentralized protocol (see \hyperref[sec:DecentralizationLN]{section~\ref{sec:DecentralizationLN}: Decentralization}), 
and anonymously (see \hyperref[sec:PrivacyLN]{section~\ref{sec:PrivacyLN}: Privacy}), intrinsic characteristics of the blockchain.
Although Lightning Network significantly increases the number of transactions per second, two scenarios, not convenient for using this Layer 2 solution, are worthy of consideration:
\begin{enumerate}
    \item The first scenario involves payments to route in the network. In case even just a single node (within the routing path) has not enough funds in the channel, the requested payment will fail. This event would require the sender to seek another route towards the recipient, and the whole process may take longer than expected\footnote{Notice that a route leading to a payment failure for the reason described above cannot be ``excluded'' a priori and simply not proposed, because of privacy of the state channels (for this purpose, see \hyperref[sec:PrivacyLN]{section~\ref{sec:PrivacyLN}: Privacy})}
    \item The second scenario which can slow down might occur in the case of uncooperative users. In particular, at the moment of channel closure, the presence of an uncooperative user (that voluntarily decides not to countersign the transaction) implies that the one who broadcasted the transaction has to wait for the timelock to expire before being allowed to spend the funds
\end{enumerate}
An important clarification is that for the first scenario, a partial solution might be to divide large transactions into many small ones (\textit{e.g.}, 1 BTC transaction might be split into ten smaller transactions of 0.1 BTC each). This paradigm will help to more likely find ways with the required amount of money.
On the other hand, the second problem described has no remedy since it is a consequence of a high-security level granted by the Lightning Network.

\subsection{Security}
\label{sec:SecurityLN}
Hash TimeLock Contract is a mechanism to remedy any uncooperative behavior in payment channels. A HTLC consists of two parts: hash verification (based on \textit{hashlock} mechanism) and time expiration verification (based on \textit{timelock} mechanism). LN uses hash locks and timelocks to ensure payment security.  \begin{quotation}A \textbf{\textit{hashlock}} is a condition placed on a transaction dictating that funds can only be spent by proving to know a secret. The sender hashes a secret (any Byte combination can serve as a secret) and includes the hash in the locking script. The receiver can spend it only if he/she can provide the original data (the secret) that matches the hash. Note that the only way he/she can provide that data is if the sender gives it to him/her, but it is a very unrealistic situation: therefore, only the person who knows the secret that was hashed will be able to use the payment. A \textbf{\textit{timelock}} is a condition that prevents from spending funds before a certain time. Timelocks require the production of a verifiable digital signature before a certain time.\footnote{This paragraph is quoted, with just some modifications, from \url{https://academy.binance.com/en/articles/what-is-lightning-network}}\end{quotation}
\par The idea behind HTLCs is that the receiver of a payment acknowledges receiving the transaction before a specific time by generating cryptographic proof of payment. After that time, the receiver cannot claim the payment anymore, returning it to the sender.\\ To better clarify the idea, suppose that Alice has to give 1 BTC to Bob; therefore, Bob will be the one to initiate the process with a payment request. The process can be summed up in three main points:
\begin{enumerate}
    \item Bob generates a \textit{payment\_secret} and keeps it to himself
    \item Bob produces the \textit{payment\_hash} by hashing the \textit{payment\_secret} and sends it to Alice 
    \item Alice creates a new commitment transaction that can be spent either by Bob if he can provide the \textit{payment\_secret} within a specific time (defined by the timelock) or by Alice if Bob has not provided the \textit{payment\_secret} by the timelock expiration.
\end{enumerate}

\par For this mechanism to work correctly, the parties exchange the hashes of their secrets when the channel is opened. For each new transaction, the participants exchange the old secrets in plaintext and the hashes of the new ones. 
\par HTLC allows overcoming the impasse problem described in one of the previous sections, in which a user decides to be uncooperative. At any moment, either participant can decide to sign and broadcast a transaction to the main chain. While the counterparty can spend the funds immediately, the participant, who broadcasted the transaction, must wait for the timelock to expire to spend the funds.\footnote{Note that this problem arises only in an uncooperative case: if both participants sign the transaction, they can both spend immediately.}
\par Here is a simple and concise example: considering again the 2-of-2 scheme with Alice and Bob, in which each participant put 5 BTC, we can assume that after a certain number of transactions, Alice has 9 BTC and Bob has 1 BTC on his side. Supposing Alice wants to redeem her funds, but Bob does not cooperate, she can anyway access her funds by broadcasting the last transaction to the main chain and waiting for the timelock. 
\par A significant characteristic of the LN added by the use of HTLC is that it naturally \textit{prevents cheating}. In fact, up to this point of the description, users can potentially broadcast an old transaction. However, secrets of the older transactions have been shared between the parties. Therefore, LN prevents these behaviors since the counterparty knows the secrets used for older transactions and can exploit them to get all the funds.
\par Despite everything, there is still a minimal possibility in which a malicious user can take advantage of the counterparty's offline period, broadcasting an old transaction without the other part being able to do anything within the timelock deadline (as he/she is offline). A significant clarification is that the vulnerability just explained has a very low percentage of occurrence, since the whole system is not managed directly by the parties but at a lower level of the network. \par For what regards Raiden Network, instead, the last described problem can be mitigated: in fact, a node going offline can ask and rely on a Monitoring Service, which is basically a (set of) node(s) monitoring the channel and reacting to possible closures of the latter; in particular, the user pays a reward to the first node which discloses a cheating behaviour by his/her counterparty, while the user himself/herself is offline. Of course, the introduction of Monitoring Service solves one of the security problems of state channels L2, at the price of losing something in terms of privacy of Raiden Network.

\subsection{Decentralization}
\label{sec:DecentralizationLN}
Given its \textit{peer-to-peer} architecture, Lightning Network can certainly be considered as a decentralized framework, at least from a theoretical perspective. In fact, no entity or node manages or regulates \textit{off-chain} transactions, which are directly performed over the \textit{peer} network in a single-hop or multi-hop way. 
\par However, even if decentralization is fully guaranteed under the theoretical point of view, the real structure of the network actually has scale-free properties: the majority of LN nodes has active channels (of limited capacity) with few \textit{peers}, while a limited number of \textit{hubs} is connected to a high number of nodes. It is interesting to understand that this \textit{hub and spoke} architecture is quite likely to be ``fed'' by single users' decisions: suppose, for instance, that Alice (A) and Bob (B) enter Lightning Network in order to establish a payment channel; their idea is that of starting transacting between themselves, but also becoming active users of LN by later exchanging cryptocurrencies with other users, too. The structure of LN, for this example, is imagined to be the following:
\begin{figure}[H]
\centering
\includegraphics[width=0.7\textwidth]{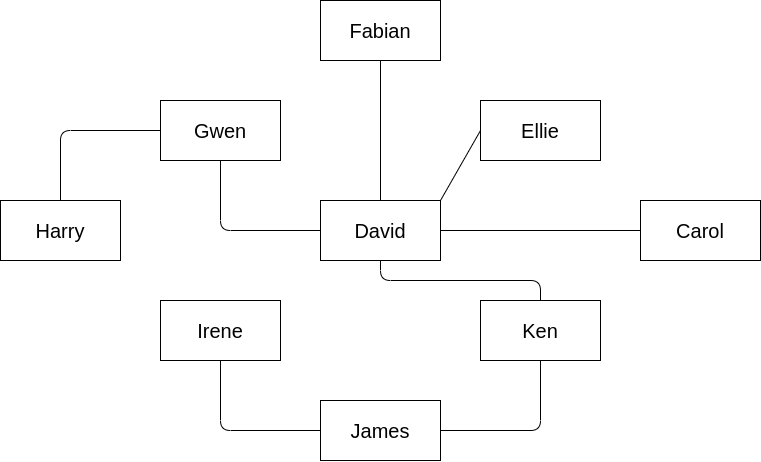}
\end{figure}
\noindent Alice and Bob have basically two choices: creating a direct channel between themselves or creating an indirect one. Of course, a direct channel would ensure no fees for all the transactions between A and B, except for the fees payed on the main chain; on the other hand, if any of the two also liked to transact with Carol (C), Gwen (G) and James (J), he/she would necessarily have to open other channels. The indirect channel, instead, would add LN fees for transactions between Alice and Bob, but would require no other channel for transacting with C, G, J.
\begin{figure}[H]
\centering
\includegraphics[width=0.68\textwidth]{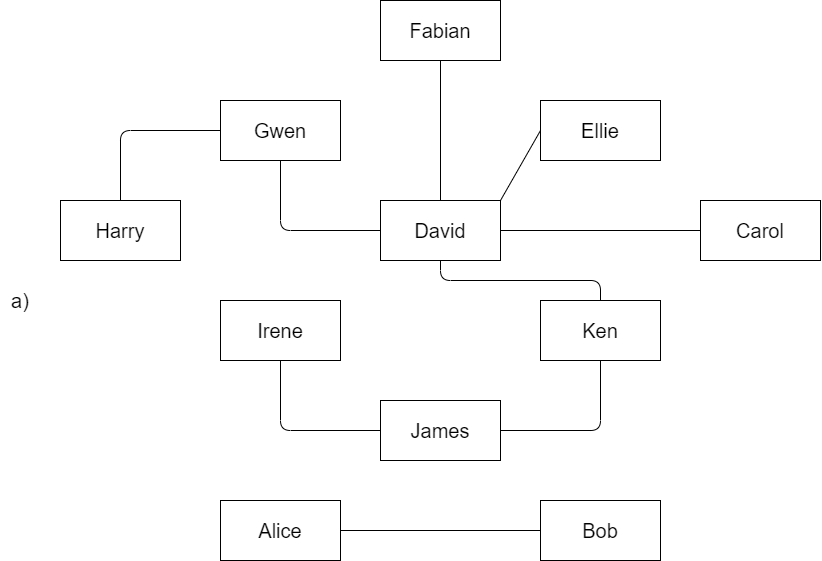}
\end{figure}
\begin{figure}[H]
\centering
\includegraphics[width=0.7\textwidth]{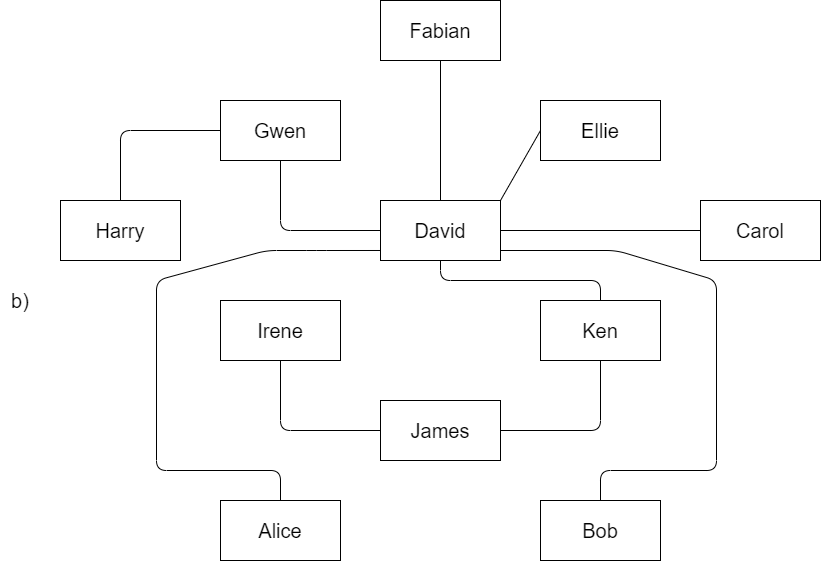}
\caption*{\textbf{Figure:} LN topology after the creation of: a) direct channel between Alice and Bob; b) indirect channel between Alice and Bob, with \textit{hub} David as intermediate node}
\end{figure}
\noindent Since \textit{off-chain} fees are sensibly lower than main chain fees, and given the fact that many nodes are likely to enter LN and keep channels active in order to transact with more than one other \textit{peer}, most of them decide to open a state channel with a \textit{hub} (instead of multiple ones with ``target'' nodes); in the previous example, at least one between Alice and Bob (if not both) are likely to open their channel with David. In reality, the presented concept is just a general idea, because also channels' capacities play a key role in the evolution of LN architecture; in any case, several studies have shown that scale-free properties have been emerging in Lightning Network, and that users tend to prefer setting channels with reliable and well-established central nodes, so to pay the smallest amount of fees (few hops with approximately every other \textit{peer}, without the necessity to open various channels, which has a considerable \textit{on-chain} cost) and be connected with the majority of nodes.
\par So, Lightning Network certainly has the idea of preserving decentralization as its core: in fact, it is implemented as a \textit{peer-to-peer} network. However, the real structure of this L2 solution tends to be less distributed than one might think. This aspect can not be considered an unbalanced step towards centralization of the blockchain, since, in any case, the architecture is still P2P and final summaries of transactions are reported \textit{on-chain}. However, it is also important to take into consideration the real structure of LN, in order to better understand possible issues for robustness (in a scale-free network, if one or more \textit{hubs} are offline there might be problems, since they constitute the ``core'' for connectivity), fee policy (as explained later, in \hyperref[sec:FeesAndMicropaymentsLN]{section~\ref{sec:FeesAndMicropaymentsLN}: Fees and micropayments}) and privacy (see \hyperref[sec:PrivacyLN]{section~\ref{sec:PrivacyLN}: Privacy}). Moreover, another aspect should be underlined: \textit{hubs} are not only central nodes, but tend to be also the only nodes which can route substantially high-value payments (in fact, most of users have only low capacity active channels); this fact is not a huge issue since LN is mainly a network of micropayments (low fees make it possible to spread a transaction into a set of smaller ones), but it is certainly another considerable part in the overall analysis.

\subsection{Privacy}
\label{sec:PrivacyLN}
Privacy of state channels Layer 2 solutions, such as Lightning Network, has to be evaluated under two different perspectives: the first one is related to confidentiality about transactions between two nodes, as well as their balances in active channels in which they are involved; the second, instead, is strictly connected to routing, which is an essential part of the technology. 
\par LN offers to users the possibility to transact for an indefinite number of times, by reporting on the blockchain only an opening and a final transaction: this implies a high level of privacy between transacting nodes, since the only information which is made public is the balance of the channel at ``time 0'' and at closing time. In practice, under this point of view, state channels guarantee more privacy than the main chain. 
\par In order to make routing possible, routing protocol over Lightning Network shares IP addresses of nodes and capacities of active channels: it is useful to remember that the capacity of a channel is the (fixed) amount of cryptocurrencies locked in the multisignature address, while balances summarize the (dynamic) evolution of funds between the two nodes. The most important information to be hidden is certainly the balance of a channel; in fact, nodes involved in an active channel and the capacity of the latter are on the blockchain since the setting of the channel itself. Hiding balances is Lightning Network's key feature for confidentiality of state channels, although it creates a bit of inefficiency in routing payments: in fact, given the impossibility to know whether all edges in a route have enough funds to support the transaction, more than one attempt might be necessary to correctly finalize a payment. 
\par For what regards multi-hop payments, another aspect ensuring confidentiality is an \textit{onion-routing} protocol, for which the sender creates several cryptographic layers which allow every intermediate node to be aware only of the identity of the predecessor and the successor in the path. 
\par In order to summarize, transactions on direct channels are completely private, while transactions with intermediaries are quite confidential; in the latter case, the longer the routing path, the higher the privacy degree. It is important to notice that a transaction involving only one intermediate node is completely not private in a certain sense, because the routing node is fully aware of the payment's size and direction; however, given \textit{onion-routing} approach, this node does not know whether the payment starts from its predecessor and is destined to its direct successor, or it is a longer-path transaction. This consideration is certainly valid, but it can be in some sense threatened by the scale-free topology of LN; in particular, most of multi-hop payments are directly routed across a limited number of \textit{hubs}, so the general tendency is to have short paths for routing transactions (and short paths certainly mean a lower level of privacy with respect to longer ones). 
\par Anyway, privacy guaranteed by Lightning Network is quite robust, especially for transactions on direct channels or long routing paths. Some issues may arise, of course, but they are caused by the topology of the network, not by the ``privacy protocol'' itself. \par For what regards Raiden Network, instead, it is worthy repeating that a bit of privacy is ``sacrificed'' in order to implement a Monitoring Service, which guarantees a higher degree of security (see \hyperref[sec:SecurityLN]{section~\ref{sec:SecurityLN}: Security}).

\subsection{Fees and micropayments}
\label{sec:FeesAndMicropaymentsLN}
As anticipated in the previous sections, Lightning Network transactions in a direct channel are not subject to fees, while intermediate nodes that route indirect payments generally ask for them (even if they are not mandatory). The main reason why LN fees are null or sensibly lower with respect to \textit{on-chain} fees is that almost negligible computational effort is necessary to process a transaction on state channels; small fees present in the network, in fact, are simply related to the fact that, for acting as an intermediate router, a node should unbalance two of its channels: in order to accept this fact, an incentive is asked. 
\par Since LN fees exist only in multi-hop payments, and routing paths depend on the topology of the network, the fee market has been evolving in accordance with Lightning Network's structure. Up to the moment of writing, the evolution of fees is still in an early phase and, as happened to many blockchains in their first times, the additional cost for transacting is very low. However, even if there is no reason connected to computational effort or scalability that forces fee price to rise, Lightning Network's free-scale topology might lead to an evolution. 
\par LN fee market is driven by two main sources:
\begin{itemize}
\item Every node acting as an intermediate router sets the fees it is going to apply: typically, fees are composed of a variable part (proportional to the size of the payment to be routed, since the higher the payment, the bigger the unbalancing factor for intermediate node) and a fixed part
\item Senders of multi-hop payments can decide which routing path to follow
\end{itemize}
In practice, senders are likely to prefer routing paths that charge the lowest fees; these paths are usually among the shortest available to reach the recipients, since every intermediate node may ask for fees. In some sense, short paths are better than long paths for what regards fees, while they are worse in terms of privacy.
\par The important aspect to consider is that, in the case in which Lightning Network \textit{hubs} significantly centralize routing payments, they could decide to charge higher fees, and this can be a problem for micropayments. In fact, suppose Alice is not directly connected to Bob in LN, and needs to send a single micropayment to him. Alice's options are either transacting \textit{on-chain}, or opening a LN channel with Bob, or routing a payment to him: of course, neither the first nor the second possibility are convenient, since they would involve at least one transaction on the main chain, with related fees; so, the only option for Alice is to route the transaction. Imagine now that Lightning Network topology is quite centralized, and at least one \textit{hub} should be traversed in order for the payment to be routed: in this situation, since Alice's only option is transacting \textit{off-chain} without opening new channels, she is going to accept not only very low fees (as the ones present in LN at the moment of writing), but a bit higher ones. This example shows that, dependently of the topological evolution of the network, fee market may evolve in one direction or another. Up to most recent data, fees in LN are low, and this is completely in accordance with the idea of this L2 solution: offering state channels to transact, which imply no additional cost under the theoretical perspective. As for privacy's situation, also for fees the most important source of issues in the future may arise from the free-scale structure of the network.

\newpage
\section{Plasma}\label{sec:Plasma}
Plasma is a Layer 2 blockchain scalability solution for Ethereum proposed by Joseph Poon and Vitalik Buterin in 2017. Plasma aims at extending the concept of sidechains, as a way to reduce the number of transactions to be processed by the L1 blockchain, also going to reduce transaction latency and cost.
\par Standard sidechains allow the deposit of assets on a contract located in the L1 chain (e.g. a smart contract on the Ethereum blockchain), which will be monitored by the sidechain operator, to be able to credit the assets themselves to users in the sidechain. Greater scalability is achieved thanks to the different consensus mechanisms, such as, for example, proof-of-authority, which allows for faster block times and thus a dramatic increase in transaction throughput, with very low fees (economic incentives might mainly derive from the service the operator is willing to offer). 
\par However, as stated in the well-known scalability trilemma, these advantages come at a cost in terms of centralization and lack of security: as an example, in a proof-of-authority sidechain, the set of validators, which are allowed to generate new blocks, could stop producing new blocks and stop processing exit requests from the sidechain (i.e. not allowing anyone to withdraw funds previously deposited in the sidechain itself). In other words, standard sidechains imply the need for trust in the chain operator.
\par Plasma proposal's goal is to solve this problem by publishing each sidechain’s block header to the L1 chain (the root chain), therefore minimizing trust but at the same time allowing verifiable fraud proofs and enforcible state, given that the root chain is secure and constantly available. This could lead to move a huge amount of load in terms of transactions, from the root chain (L1) to Plasma chains (L2), with only periodic commitments on the root chain. Moreover, Plasma reduces the storage needed in the root chain, by compacting several state transitions in a single merkle root commitment.
\\
\begin{figure}[H]
\centering
\includegraphics[width=0.7\textwidth]{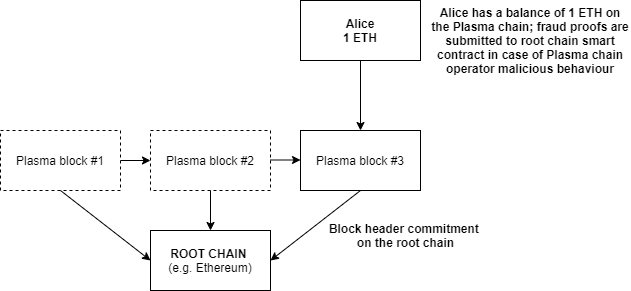}
\end{figure}

\par Ideally, multiple Plasma chains can be created, each one with a different purpose and use case. As an example, one could create a Plasma chain to handle transfers involved in a decentralized exchange application, while another Plasma chain could be used to handle micropayments. In addition to this, child chains could be organized as a tree of chains, in a court system fashion, where the chain at height one of the tree can solve disputes arising from any height two chain, as it holds all blocks headers and can evaluate proof of frauds. The proposed design resembles the flow of the \textit{MapReduce} computation paradigm, by assigning workloads to child chains (at various depth of the tree), which then commit work up to the root chain.

\begin{figure}[H]
\centering
\includegraphics[width=0.7\textwidth]{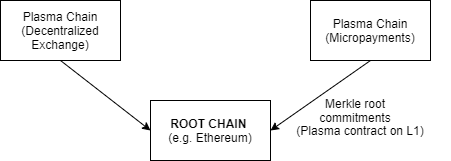}
\end{figure}

\par The described construction allows user to actively participate only if their \textit{on-chain} balances are being updated or some malicious behaviour on the Plasma chain is being detected. Funds can be held in the child chain, and they are secured by a full representation on the root chain: assets can then be withdrawn after a dispute mediation period, where others can submit fraud proofs. Fraud proofs can stop a withdrawal process if irregularities are detected, for example if the coins are proven to have already been spent. Penalizations are put in place, by appending a bounty alongside withdraws and proof of frauds which is lost in case of faulty behaviour.
\par To achieve this verification scheme, an UTXO (Unspent Transaction Output) model is adopted for the Plasma chain (at least in the initial proposal), because it allows more compact ways to verify if a particular state has been spent, and, as a result, it is more efficient for fraud proofs and withdrawals. This idea is in contrast with Ethereum account-based state model: in UTXO chains there is no concept of balance, but there is a set, a data structure that is updated every time a new transaction is sent. Every transaction consumes an output in the previous set, which was fed by a previous input; the new output produced becomes an input for the next generated element of the set. In this type of chain, the final balance for a given address is reconstructed by summing all the unspent outputs for that address. The UTXO model exploited by most Plasma chain implementations causes general computations (e.g. smart contracts execution) not to be supported; only simple transactions such as transfers or swaps are possible on the majority of Plasma variants.

\subsubsection*{Fraud Proofs}
\label{sec:FraudProofs}
Plasma child chains are operated by validators, which propose blocks. Plasma's enforcible state restricts their possible malicious behaviour by making use of fraud proofs; in case a malicious block is propagated, any other actor that is monitoring the child chain and receives the block can submit a merkleized fraud proof on the parent blockchain. The invalid block is then rolled back, and the proposer of the faulty block is penalized.\footnote{All images in this section have been inspired by \url{https://plasma.io/plasma-deprecated.pdf}}\begin{figure}[H]
\centering
\includegraphics[width=0.7\textwidth]{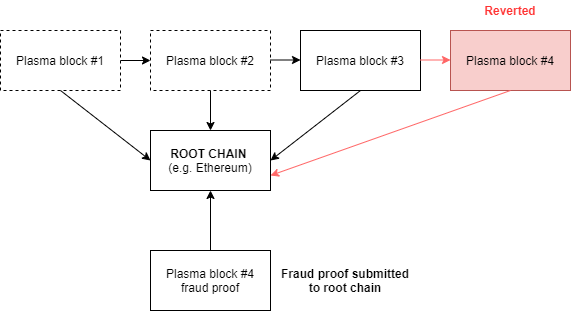}
\end{figure}
Both fraud proofs and withdrawals are secured by the concept of merkle proof. The information that is committed by the Plasma chain to the root/parent chain is the merkle root of the block that has just been generated. Whenever a user wants to perform a withdrawal, a request is submitted to the root chain contract, together with a merkle proof, that is checked by the contract: these checks consist of making sure that the transaction which generated the output that the user is willing to withdraw was included in a certain block in the past; moreover, the contract checks if the specific output \textit{belongs} to the user who is performing the request. 
\par However, the user cannot prove that that output has not already been spent within the Plasma chain; therefore, a challenge period is needed in order to allow other actors to possibly provide proofs (still in the form of merkle proofs) showing that the output cannot be redeemed.
\par For this mechanism to work, block data availability is needed, in order to be able to produce fraud proofs.
It is important to note that, after some time and if the parent blockchain reaches sufficient finality, commitments are considered finalized and cannot be reordered.

\subsubsection*{Deposits and Withdrawals}
Plasma's \textit{on-chain} architecture is composed of contracts that have the following responsibilities:
\begin{itemize}
\item Keeping track of block header hashes (merkle roots) commitments sent by Plasma child chains
\item Processing submitted fraud proofs and withdrawals, and handling the related penalization for the invalid ones
\end{itemize}

\par Deposits are initiated by the user by sending funds to the contract located in the root chain. The Plasma chain detects that a deposit has been started, and includes in a block a commitment to the fact that funds will be spendable. Finally, the depositor signs a transaction on the Plasma chain, to declare that he/she has seen the commitment made by the chain in a previous block. \\

\par \noindent The withdrawal process is composed as follows:
\begin{enumerate}
  \item The user sends a signed withdrawal transaction on the root/parent chain. The withdrawn amount must correspond to the whole unspent output. An additional bond amount is put at stake to allow for the penalization of invalid withdrawal attempts
  
  \item There is a predefined timeout period to allow for disputes. In case valid proof is submitted, the withdrawal is canceled and the bond staked before is lost (this is useful to discourage faulty withdrawal attempts). If no fraud proofs are found, then the user is allowed to redeem the requested funds
\end{enumerate} 

\par \noindent Withdrawals as described above are quite slow in nature, due to the dispute period needed to allow others to provide fraud proofs. Usually, a period of time considered adequate as challenge period is 7 days. This mechanism protects users from frauds, but also comes at a great cost in terms of usability and user experience.
\par Fast withdrawals are possible too, exploiting liquidity providers (LP) willing to offer this as a service. If the liquidity provider is able to fully validate the Plasma blockchain, then user funds could be locked to a contract on the child chain. The behaviour enforced by the contract is that, if a certain transaction proof stating that the LP has sent funds to the user can be found on the root blockchain and has been finalized, then the payment to the LP will go through in the Plasma chain; otherwise, the sender can redeem back his funds. Liquidity providers are incentivized to provide this service by charging extra fees that the user has accepted to pay in order to have a faster way to exit the Plasma chain (with respect to a simple withdrawal). 
\par Basically, this mechanism implements an atomic cross-chain swap, and can also be exploited to move funds between Plasma chains.

\subsection{Scalability}
\label{sec:ScalabilityPlasma}
In terms of scalability, a Plasma chain can claim the same performance improvements achieved by standard sidechains. Obviously, the transactions throughput and latency that can be obtained ultimately depends on the consensus mechanism that the Plasma chain exploits. Usually, we can expect child chains to use a consensus algorithm other than proof-of-work, such as proof-of-stake (PoS) or proof-of-authority (PoA), which allow for faster block times and a higher number of transactions per second (TPS). 
\par As an example, \textit{Polygon} (formerly known as \textit{Matic Network}), a commercial adapted implementation of Plasma exploiting PoS as consensus mechanism for the Plasma sidechain, can be considered. Polygon claims to have reached a peek of 7,200 TPS on its internal testnet; at the moment of writing, Polygon mainnet has an average block time of 2.1 seconds and a gas limit per block of 20 million. Using current available data, simple statistics can be computed, in order to get the current average gas consumed for each Ethereum mainnet transaction: 
$$Txs\approx1,500,000\;transactions\;per\;day,\;Blocks\approx6,500\;blocks\;per\;day$$
$$Average\;transactions\;per\;block=\frac{Txs}{Blocks}\approx230\;transactions$$
Considering the current Ethereum mainnet block gas limit (12,500,000):
$$Average\;gas\;per\;transaction\approx\frac{12,500,000}{230}\approx54,350$$
As a consequence we can obtain an approximate value for the current Polygon throughput:
$$Throughput\approx\frac{20,000,000}{Average\;gas\;per\;transaction}\cdot\frac{1}{2.1\;seconds}\approx175\;TPS$$
The obtained result is a good improvement (about 11 times) over Ethereum mainnet current thoughput, but still far from what is needed in order to envision mass adoption of the technology and handle, for example, the load implied by global retail transactions. However, it is interesting to notice that the approximation adopted here is quite conservative: blockchains with similar consensus mechanisms (not necessarily implementing Plasma) exploit a greater block gas limit and, potentially, a lower block time: this could lead to further improvements in terms of throughput. Moreover, this type of construction implies that other additional Plasma sidechains can always be added, therefore enabling \textit{horizontal scaling} of blockchains.

\subsection{Security}
Security on a Plasma chain is ultimately achieved by merkle proofs submitted to the root chain, which prevent child chain's validator Byzantine behaviours and invalid attempts to withdraw outputs (such as those already spent). 
\par Most of the security concerns in this context are related to data availability and block withholding attacks (BWA), which happen when validators refuse to publish new blocks. Users of a Plasma chain must constantly monitor the chain and, in case of BWA, they need to perform a \textit{mass exit}. In addition to the fact that the need to constantly monitor the child chain (\textit{liveness assumption}) creates usability issues, one of the main challenges in Plasma is to make mass exits compact and efficient. Mass exits cannot be performed as \textit{fast withdrawals} since, in this case, the chain is assumed to be Byzantine and liquidity providers would not have the interest in process this kind of withdrawals. Even if it is theoretically possible to use the simple withdrawal process described in the previous section, a dedicated procedure has to be designed to mitigate data availability issues as much as possible.
\par The mass withdrawal interactive game proposed by Buterin and Poon is based on a mass exit operator, which verifies the Plasma chain up to the point in which the BWA started. An exit transaction is created with an attached massive bond and users sign of the mass withdrawal. The operator will then sign and broadcast the \textit{mass exit initiation transaction} (MEIT) to the destination chain, optionally charging a fee to every user who is participating in the process. The operator also publishes a full \textit{bitmap} of the state involved in the exit: this let observers of the Plasma chain to understand what is being withdrawn and optionally challenge it; finalization of the MEIT may take a lot of time, in the order of weeks. 
\par However, the described design leads to problems that are intrinsic to mass-exits: supposing a very popular Plasma chain with a very big number of users, a fraudulent event by validators could cause the root chain to be flooded by huge simultaneous mass-exit transactions. As a result, the L1 chain would likely be heavily congested. \\

\par Other possible security threats may include:
\begin{itemize}
    \item Smart Contract code vulnerabilities: smart contracts on the root chain are responsible to enforce the state of the child Plasma chains. The security of the Plasma architecture is entirely reliant on complex \textit{on-chain} smart contracts; the complexity needed in this context increases risks of vulnerabilities
    \item Redeem transactions may be too expensive on the root chain, thus making withdrawals unfeasible
    \item State enforcement and execution of exit transactions might fail if a 51\% attack is being performed on the root chain and blocks are being censored
\end{itemize}

\subsection{Decentralization}
\label{sec:DecentralizationPlasma}
From a strictly theoretical point of view, the level of decentralization offered by a Plasma chain exclusively depends on the exploited consensus mechanism of the chain itself. However, Plasma is by construction trying to let not fully decentralized sidechains (such as those using proof-of-authority) to avoid the usual problems deriving from a centralized chain, through the use of merkle proof commitments on the root chain. In practice, the effects of the decentralization level chosen for a Plasma chain depend on two factors:
\begin{itemize}
    \item As previously stated, the consensus mechanisms of the Plasma chain
    \item \textit{When} transactions are considered finalized
\end{itemize}
To avoid side effects deriving from an increased level of centralization, one should consider a transaction finalized only when the merkle root of the block containing the transaction itself is committed to the root chain: doing this ensures that the state is being enforced by fraud proofs, and that the security is the same as L1 blockchain. To mitigate the need for waiting the commitment, the Plasma chain could make use of a proof-of-stake consensus mechanism, in order to find a balance between decentralization and achieved scalability: for this reason, \textit{Plasma Proof of Stake} was proposed by Buterin and Poon, also trying to reduce the risk of block withholding attacks.

\subsection{Privacy}
Although privacy issues are not a primary concern in Plasma design, some considerations regarding the consequences of the proposed solution can be done. Essentially, Plasma is based on the concept of avoiding broadcasting all transactions in the root ledger. The load on the L1 chain is reduced, and the resulting privacy is determined by the specific child chain design, which could also be focused on guaranteeing greater privacy properties with respect to public ledgers like Ethereum or Bitcoin. Theoretically, implementations for such a Plasma chain could be inspired by already existing designs which aim to provide an increased level of privacy, such as \textit{ZCash}. A solution of this kind would widen the scope of this Layer 2 solution, allowing not only for scalable public blockchains, but also to add privacy features to already existing public decentralized ledgers.

\subsection{Fees and micropayments}
In short, fees on the root chain are useful to increase security and avoid spam attacks, and to serve as an incentive for validators or miners to behave correctly, especially in proof-of-work blockchains. 
\par In Plasma sidechains, transactions fees can be reduced dramatically. Disincentives for extremely low value transactions and spam attacks are still a requirement, but there is no more the need for a particularly decentralized consensus algorithm, since users can always exit and return to the secure and decentralized root chain. As a consequence, a child chain can be designed, for instance, with a proof-of-authority consensus; this fact means that there are no longer miners interested in the ``race'' for high fees (situation in which miners prioritize transactions with higher fees, causing many blockchain's use cases to become unfeasible due to transactions costs). So, it is clear that Layer 2 solutions like Plasma enable features for blockchains that, in practice, are not possible on current implementations of L1 chains.
\par As an instance, \textit{micropayments} are one the new possibilities enabled by Layer 2 scaling. One possible design choice is to create a Plasma chain specific to offer micropayments. A solution like this would allow real world scenarios like coffee shops using Plasma to receive payments by customers: the shop would need to monitor the child chain to detect malicious validators behaviour and periodically pay a fee on the root chain to withdraw earnings, for example once a week. Micropayments are considered a well-suited Plasma use case also for the resource saving in terms of storage which the root chain could benefit from (we assume that very low value transfers do not need the level of security offered by L1 chains), as a great number of transactions can be compressed in a light block header. In addition, another factor that favors this kind of low value transfers is the fact that, in a Plasma chain, it is possible to send assets to users who are not currently in the set of participants, unlike classical state channels solutions.

\subsubsection*{Plasma variants}
Since 2017 many different designs, derived from the above described Plasma specification, have been proposed. The most significant ones are briefly described below.

\begin{itemize}
    \item Plasma MVP: Minimum Viable Plasma is a proposed simple UTXO-based Plasma chain. It targets high volumes of payments, but does not support general computation (smart contracts). MVP's consensus mechanisms is proof-of-authority, thus it can also be used for private blockchains.
    \item Plasma Cash: user funds deposited to the Plasma chain are represented by non-fungible tokens (NFTs). Consensus is typically implemented as proof-of-authority or proof-of-stake.
    Thanks to the fact that Plasma Cash makes use of \textit{sparse merkle trees} instead of standard ones, the solution is highly scalable, but suffers handling applications that need to deal with fraction of assets (an example include micropayments). As a result, this variant seems to be appropriate to directly handle NFTs \textit{off-chain}, which is an extremely useful feature for supply chains or card games applications.
    \item Plasma Debit: it aims to mitigate problems deriving from Plasma Cash by making use of payment channels inside the Plasma Chain, with operators as \textit{routers} between users involved in the transfer. Therefore, any user must have payment channels set up with the chain operators. By using channels, this solution allows for transfers of fractional parts of funds.
\end{itemize}

\par These new versions of Plasma would undoubtedly deserve a detailed discussion, which (however) is outside the scope of this document.

\section{Rollups}\label{sec:Rollups}
Payment channels and Plasma are usually considered ``\textit{full}'' Layer 2 solutions, while \textit{Rollups}, a framework for Ethereum proposed in 2018, is considered a ``\textit{hybrid}'' solution between L1 and L2 scaling: this is due to the fact that, while Lightning Network and Plasma summarize many state transitions in one single commitment or channel closing, the Rollup proposal implies that some information on every single transaction sent on the L2 is posted \textit{on-chain}. The idea of Rollup is that transactions are aggregated \textit{off-chain}, causing a reduction in congestion and fees on the root chain.
\par The concept of Rollups is somehow similar to the one previosly explained for Plasma: a smart contract deployed on the root chain keeps track of the current (most recent) merkle root of the state of the \textit{rollup} (in practice, this could be the L2 chain). The computation of the new state is performed \textit{off-chain}, and the root chain is used for data availability (by posting a piece of data for each transaction). The described design allows to avoid data withholding issues, because all the information is always retrievable from the root chain, which is considered to be secure and always available.
Any actor is able to publish a \textit{batch}, which is a collection of compressed transactions with the previous merkle root and the new merkle root attached to it. At this point, the Rollup contract, before updating the state root with the new one, has to check (for security reasons) if the previous root corresponds to the current root (for a focus in the security analysis of the solution, refer to next discussions in the section). A peculiar feature of Rollup is that it allows to transact outside the Rollup contract itself: this is meant to support transactions whose input comes from outside or whose output is destined for outside. At the time of writing, there are mainly two types of Rollup solutions: ZK Rollups and Optimistic Rollups. They differ in how they perform verification on the submitted batch.

\subsubsection*{ZK Rollups}
The ZK Rollups solution is based on the concept of \textit{validity proof} and \textit{zero-knowledge proof}. The idea behind ZK Rollup is to bundle every batch with proof of their validity. The proof should be easy to check, proving the correctness of the batch content. ZK Rollup framework uses the SNARK proof (see \hyperref[sec:SecurityRollup]{section~\ref{sec:SecurityRollup}: \textit{Security}}): this kind of system allows observers to immediately prove the validity of an assertion. So, the main characteristic of this proof is that it is cheap to verify. However, computing it is expensive. Therefore, ZK Rollup is an appropriate system for transactions' management, but it does not fit with complex contracts execution.

\subsubsection*{Optimistic Rollups}
Optimistic Rollups is an \textit{interactive} approach. It is, in a sense, more similar to Plasma's, since it involves the use of \textit{fraud proofs} (see \hyperref[sec:FraudProofs]{\textit{Fraud Proofs} in Section 5}). In this kind of solution, new batches (and, therefore, new merkle roots) are published by operators, without being proved to be right by the Rollup smart contract. This mechanism is further detailed in \hyperref[sec:SecurityRollup]{section~\ref{sec:SecurityRollup}: \textit{Security}}.

\subsection{Scalability}
\label{sec:ScalabilityRollup}
One of the key points of Rollups scaling is compression: in practice, every Rollup transaction (remember that information of every transaction is reported \textit{on-chain}) may be compressed to occupy a total space of (down to) approximately 12 Bytes. This size, of course, varies with Rollup and transfer types: Optimistic Rollup transactions need more Bytes (information) for later verification (due to the absence of a SNARK proof); ETH transfers take less Bytes than more ``complicated'' transfers, as it normally happens on Ethereum blockchain. So, 12 Bytes lower-bound for a Rollup transaction refers to an ETH transfer performed using ZK Rollups. \par To better understand the translation of this compression feature into throughput achievements, a best-case scenario can be considered: ZK Rollups for transferring ETH. Considering the current values, on Ethereum blockchain, of:
\begin{itemize}
\item Gas limit: 12.5 Million gas/block
\item Gas per Byte in L1 transaction: 16 gas/Byte
\item Average block time: 13 sec/block
\end{itemize}
Supposing to spend 1 Million gas for proof verification, the following calculations can be developed:
$$Block\;size=\frac{Gas\;limit - Gas\;for\;proof}{Gas\;per\;Byte}=\frac{(12.5 - 1)\;Million\;gas/block}{16\;gas/Byte}\approx715,000\;Byte/block$$
$$Transactions\;per\;block=\frac{Block\;size}{Rollup\;transaction\;size}\approx\frac{715,000\;Byte/block}{12\;Byte/tx}\approx59,500\;tx/block$$
$$Throughput=\frac{Transactions\;per\;block}{Average\;block\;time}\approx\frac{59,500\;tx/block}{13\;sec/block}\approx4,500\;TPS$$
It is crucial to remember that this is a theoretical achievement for ETH transfer on ZK Rollups, considering that the whole \textit{on-chain} transaction is composed of data coming from the Rollup itself. In any case, although further complications may be analyzed, 100x scalability improvement is a good approximate evaluation of ZK Rollups scalability with respect to Ethereum L1 throughput. \par For what regards Optimistic Rollups, instead, considering a Rollup transaction size of 12 Bytes is not possible anymore, since further data must be committed on the main chain for verification; however, even considering a transaction size 6 times bigger than the one for ZK Rollups (which is a quite reasonable value), the obtained throughput is approximately more than 800 TPS: so, still substantially higher than Ethereum current throughput for token transfers. At the time of writing, one of the main features offered by Optimistic Rollups with respect to ZK Rollups is \textit{general computability} (of smart contracts). In any case, also for complex contract interactions the improvement measured as a ratio with respect to L1 scalability is reasonably the same as ETH transfers, since other contracts normally require more computation on the main chain too. \par In addition to pure throughput analysis, it is important to analyze the withdrawal time: in fact, users' purpose is not only to transact at high speed, but to drop out fast too. At this purpose, the two Rollup solutions behave very differently, due to security verification diversity (see \hyperref[sec:SecurityRollup]{section~\ref{sec:SecurityRollup}: \textit{Security}}): while for ZK Rollups the withdrawal period is very fast (thanks to SNARK proof, waiting for the next batch is sufficient), in Optimistic Rollups this time goes up to 1 or 2 weeks, so to be compatible with fraud proof challenge period.

\subsection{Security}
\label{sec:SecurityRollup}
Dealing with security in the Rollup system, it is crucial to divide the discussion by distinguishing ZK Rollup and Optimistic Rollup. \par ZK Rollup is a schema that involves two kinds of users: transactors and relayers. The formers are users who are willing to perform transfers; relayers, instead, are users that have staked a bond in the Rollup smart contract and are in charge of collecting a large number of transactions to create a \textit{rollup}. ZK Rollup, as introduced in a previous section, exploits the SNARK proof system.\footnote{A detailed description of SNARK proofs goes outside the goals of this work; the interested reader can find a meaningful reference at \url{https://z.cash/technology/zksnarks/}} Relayers (that be assimilated to operators of a Plasma child chain) are in charge of generating the SNARK proof, which is a hash that represents the delta, the difference between the precedent state of the blockchain and the new state, after the execution of the transactions that compose the \textit{rollup}. \\
After transfers have been bundled up, the batch is sent to the root chain contract together with the SNARK proof, guaranteeing that the new state has been correctly generated by the transactions included in the batch itself (starting from the old merkle root up to the new one). This kind of approach can be considered as an \textit{explicit verification}, which eliminates the need for fraud proof. Since their verification complexity is $\mathcal{O}(1)$, SNARKs are well-suited for transaction managing and small contract execution applications. However, as already anticipated, this mechanism does not fit well with more complex contracts' execution due to their complicated proof generation. \par The Optimistic Rollup approach, instead, is similar to Plasma from a security point of view. Both of them implement the same proof mechanism: fraud-proof. This interactive approach allows overcoming the complex contracts execution management. The general idea is that the \textit{rollup} contract keeps track of the history of updated states, and anyone (challenger) who detects a wrong post-state root can publish proof of that incorrectness. At that point, the contract can verify the provided proof: if the proof points out an invalid submitted assertion, the system restores the state considering the last valid batch and penalizes the publisher.

\subsection{Decentralization}
In addition to what already presented in the discussion dedicated to Plasma decentralization (see \hyperref[sec:DecentralizationPlasma]{section~\ref{sec:DecentralizationPlasma}}), which can be partially extended to Rollups, a few points should be considered. \par First of all, it is important to remember that Rollups are implemented \textit{on-chain} through a smart contract: this, of course, implies a higher rate of centralization, especially if Rollups usage becomes substantial and, in consequence of this, also the volumes managed by a single smart contract. On the other side, contrary to other kinds of L2 solutions (such as state channels), in Rollups all data needed to reconstruct every state is published \textit{on-chain}, thus ensuring a consistent level of decentralization under this point of view. \par An important centralization issue concerning ZK Rollups, however, is the following: since SNARK proofs must start from a trusted initial condition, the latter can be a source of centralization for the early life of the Rollup, because it is controlled by a small group. In any case, the problems related to this aspect are closer to possible attacks mining security rather than menaces for the centralization of the L2 solution.

\subsection{Privacy}
Rollups cannot certainly be considered as a Layer 2 solution which has, among its biggest interests, privacy; in particular, by reporting on the blockchain a ``summary'' of every single transaction performed \textit{off-chain}, the privacy level is certainly lower than all the solutions committing only an initial and a final transaction. One countermeasure to this can be found, at least for Optimistic Rollups: they can act as Layer 3 scalability solution on top of an already existing L2 for Ethereum ensuring privacy of transactions; of course, this proposal comes at the cost of excessive fragmentation, which translates the ``privacy countermeasure'' for Rollups into a not sufficiently good idea. \par At the moment of writing, privacy concerns for Rollups are not considered as a primary deal, and the community is further studying and developing solutions for higher throughput, increased security and low fees much more than focusing on privacy.

\subsection{Fees and micropayments}
Fees in Rollups are composed of two parts: \textit{off-chain} and \textit{on-chain} ones. Of course, the cost associated to every transaction depends also on the type of used Rollup solution: as explained in previous sections, Optimistic and ZK Rollups differ for transaction size and verification mechanism. \par In order to present a brief example\footnote{All data considered in the presented example comes from \url{https://etherscan.io} and \url{https://zkscan.io/}} and give the flavour of the effects on fees given by using Rollups, a simple use-case can be highlighted: ZK Rollups used for cryptocurrency (ETH and ERC-20 tokens) transfers. For this purpose, the last ten batches\footnote{The decision of considering only ten batches directly derives from the scope of the example itself: providing some approximated numerical estimations as a comparison between Rollups fees and main chain ones, and not a full analysis.} (at the moment of writing) by \textit{zkSync}, an implementation of ZK Rollups, are taken into consideration. By analyzing the blocks committed and verified on the Ethereum mainnet, it turns out that each of them contains data related to between 129 and 206 transactions; the \textit{on-chain} fees associated to a block range from 0.052 Ether to 0.076 Ether, with consequent average \textit{on-chain} cost per transaction of 0.000461 ETH ($\sim$1.12 USD, at the moment of writing). These \textit{on-chain} fees are paid by validators, when a block is published on L1. However, in this case we are considering as all transactions in the blocks were transfers, but, in general, different type of \textit{off-chain} operations have different costs (e.g. withdraws are more expensive than transfers); therefore, the estimated fee must be considered as an \textit{upper bound} value for transfer fees. As a matter of fact, considering the last ten \textit{zkSync} transfers, an average fee of 0.000046841 ETH (currently $\sim$0.11 USD) was observed. It is worth to notice that this fee cost diverges from that estimated by the \textit{zkSync} team (around 0.001 USD). For a more detailed discussion, please refer to the \hyperref[sec:UseCaseCustomersUsabilityRollups]{section~\ref{sec:UseCaseCustomersUsabilityRollups}: proof of concept usability}.
\par The presented example finds its conclusions when comparing Rollups fees with Ethereum fees: for this reason, the last ten transactions (at the moment of writing) concerning transfers on the mainnet are analyzed, discovering that their average fee cost is approximately 1.23 USD. In practice, Rollups fees for cryptocurrency transactions are, in the analyzed scenario and data, up to 11 times cheaper than those performed with no L2 solution. \par Of course, one of the goals of Rollups is to decrease transaction cost more than what was inspected by the presented example. The actual values, in fact, are driven by the same reasons for which also the scalability cannot reach its theoretical achievement. In any case, a massive introduction of Layer 2 solutions would help fulfilling this goal. \par Moving to micropayments, the concept is the following: 0.11 USD (the average computed value considering current available data) still seems to be quite high considering, for instance, a transfer to pay a coffee (1 USD). This line of reasoning is true for the current situation; however, Rollups have the potential to decrease current Ethereum fees by orders of magnitude, thus enabling the possibility of payments of whatever scale.

\newpage
\section{Comparative table}
In this section, a comparative table of the analyzed solutions is presented. The table aims at bringing out the key points of each framework, by highlighting strengths and weaknesses for each of the five aspects under consideration. For a detailed content on individual solutions, please refer to \hyperref[sec:LN]{section~\ref{sec:LN} (for Lightning Network)},
\hyperref[sec:Plasma]{section~\ref{sec:Plasma} (for Plasma)} and \hyperref[sec:Rollups]{section~\ref{sec:Rollups} (for Rollups)}.\\

{\renewcommand\arraystretch{1.4} 
\begin{longtable}{|p{0.22\textwidth}|p{0.25\textwidth}|p{0.25\textwidth}|p{0.25\textwidth}|}
\cline{2-4}
\multicolumn{1}{c|}{} & \multicolumn{1}{c|}{\textbf{LN}} & \multicolumn{1}{c|}{\textbf{Plasma}} & \multicolumn{1}{c|}{\textbf{Rollups}} \\
\hline
\vspace{32mm\textbf{Scalability}} & It increases scalability by theoretically reaching around 10,000 TPS. However, there are some scenarios to consider that can delay transactions (not enough funds within the routing path) or withdraw funds (uncooperative behaviors for channel closure).
& Commercial solutions claimed to have achieved a peak of 7,200 TPS. Currently, throughput on Plasma platforms is around 175 TPS: this result has been possible by using a different consensus mechanism on the Plasma chain, in addition to a shorter block time and a higher block gas limit with respect to Ethereum.
& Thanks to compression, Rollups achieve theoretical maximum throughput limits of 4,500 TPS (ZK Rollup) and 800 TPS (Optimistic Rollup). However, practice shows that it is difficult for all optimality conditions to occur, reaching about 30\% of the theoretical throughput.\\
\hline
\multicolumn{4}{r}{\footnotesize\itshape
Continue in the next page}\\
\hline
\vspace{33mm\textbf{Security}} & 
The use of hashlock and timelock mechanisms guarantees security. HTLC overcomes the problem of uncooperative behaviors. Moreover, malicious behavior is mitigated, as the attacker risks losing all funds. &
Security is achieved through merkle root commitments to L1 that prevent operator's malicious behavior (unlike traditional sidechains). Most concerns are related to block withholding attack and the mass-exit problem, which are still open issues. &
Different system for each solution. In particular, ZK is based on the concept of \textit{validity proof}, using the SNARK system: it couples each batch with proof of its validity (easy to verify but difficult to generate). Optimistic Rollups, like Plasma, are based on fraud proofs: the idea is that anyone can provide proof of an incorrect batch which will then be verified as well.\\
\hline
\vspace{28mm\textbf{Decentralization}} & 
Fully guaranteed from a theoretical perspective. However, the real structure consists of a few active channels between users and a limited number of \textit{hubs} connected to a large number of nodes. &
Consensus mechanisms exploited in Plasma are usually less decentralized than L1, in order to achieve better performance. In practice, a trade-off between decentralization and performance must be found, for example by using PoS on L2.& 
Although on the one hand the Rollups have a high decentralization guaranteed by the main chain (albeit in summary form, all transactions are stored in the blockchain), on the other, the transactions are managed by a limited number of smart contracts, effectively increasing centralization.\\
\hline
\vspace{30mm\textbf{Privacy}} & 
High level of confidentiality: transactions on direct channels are completely private, while the level of confidentiality with intermediaries increases as the length of the path increases. & 
Not a primary concern in Plasma, but since the concept is avoiding to broadcast all the transactions to the root chain, the privacy heavily depends on the design of the Plasma child chain. & 
Surely, it is the solution that guarantees less privacy among the analyzed. Precisely, due to the fact that a summary of all transactions is published in the main chain, the transactions somehow ``remain public'' (at least in part, from which the entire transaction can be traced).\\
\hline

\vspace{26mm\textbf{Fees and micropayments}} & 
Transactions in a direct channel are not subject to fees, while intermediate nodes that route indirect payments generally ask for them as an incentive. At the moment of writing, the additional cost for transaction (multi-hop scenario) is very low. L1 fees are present for channel opening and closure. & 
Transactions on L2 are much cheaper than those on L1, because a Plasma chain can exploit a less decentralized consensus mechanism. Nevertheless, even if low, they are still needed to discourage spam attacks on the Plasma chain. L1 fees are still needed to perform deposits and withdrawals.  & 
The best case occurs with ZK Rollup: it implies negligible \textit{off-chain} fees ($\sim$0.001 USD for transfers, higher for withdrawals). \textit{On-chain} fees are still required for deposits. Optimistic Rollups have higher fees compared to ZK, even if they are still low if compared to L1.\\
\hline

\end{longtable}}

\newpage
\section{Layer 2 Use Case: Supermarket}
In order to be able to test the described Layer 2 solutions in practice, a proof of concept of a suitable use case for these technologies can be proposed. The application simulates a possible architecture put in place by a supermarket chain, to allow customers to pay using cryptocurrencies.
\par The factors under analysis for each Layer 2 option are the following;

\begin{itemize}
\item The commitment that the shop should undertake to implement the solution, both in terms of software engineering effort and costs
\item Usability for customers
\item The achieved performance (throughput and latency)
\end{itemize}

\par Basically, this sample application implies the creation of a parallel payment system with respect to pre-existing ones. This therefore falls into the case of micropayments, which have been detailed in the previous sections regarding each solution. 
\par The first of the previously described parameters also includes a brief analysis about the current state of the art of tools for a specific L2 platform. As a matter of fact, this is crucial to determine how much development effort is required by the owner of the system, and it is an indication of how mature the platform ecosystem is.
\par In this context, the term \textit{usability} is referred to a combination of two main aspects: the ease of use for customers and the cost per transaction. 
\par Regarding performance, in this case an analysis in terms of throughput (TPS) is not enough: what has to be taken into account is also the transaction latency, i.e. the time that each single user has to wait in order to have its payment confirmed by the underlying architecture. In fact, a payment method implying an average confirmation time of, for instance, 10 minutes, is not usable in a retail context.
\par It is important to point out that the results in terms of throughput that will be shown in the following sections are related to the number of transactions per second that were dedicated only to the application, not the overall performance of the tested platforms. Furthermore, the performance analysis presented in this work should be considered only as an indication of the performance of each Layer 2 solution; this is caused by the fact that \textit{benchmarking} this type of applications on public testing networks leads to results that are not always reproducible and extremely variable (mainly due to changes in the level of congestion and average fees), especially if performed using external API providers.
\par Theoretical results have been discussed in previous specific sections (\hyperref[sec:ScalabilityLN]{section~\ref{sec:ScalabilityLN} LN}, \hyperref[sec:ScalabilityPlasma]{section~\ref{sec:ScalabilityPlasma} Plasma}, \hyperref[sec:ScalabilityRollup]{section~\ref{sec:ScalabilityRollup} Rollups}). For an in-depth analysis we refer to other works designed to measure this type of statistics more thoroughly.

\newpage
\subsubsection*{Proposed Schema}
The proposed flow for the proof of concept is shown below. \\
\begin{center}
    \includegraphics[width=0.6\textwidth]{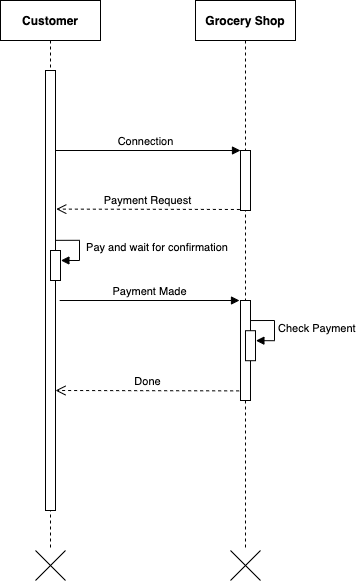}
\end{center}

\par For all Layer 2 solutions under analysis, interaction between customers and the supermarket has been emulated by using the \textit{socket.io} Node.js library to implement real-time communication. In a real-world scenario, the initial setup needed for the communication would be replaced by other systems, such as a QR code shown by the cashier to the customer; the latter would then use a mobile application released by the supermarket, which in practice would be a Layer 2 cryptocurrency wallet. This schema would allow the system to remain as decentralized as possible, because private keys are kept locally on users' devices, and customers have full control of their funds.

\subsubsection*{Common usability drawbacks}
\par Decentralization introduces some disadvantages in terms of usability that are common to all the solutions shown below: 
\begin{itemize}
    \item Users must be able to securely manage private keys and cryptocurrency wallets
    \item The need for customers to handle deposits and withdrawals, with the resulting fees \textit{on-chain} to be payed by users
\end{itemize}

\subsubsection*{Estimated performance requirements}
\par To evaluate the various frameworks exploited in the following sections, an estimated target throughput has been computed. A nationwide supermarket with the following properties was considered:
\begin{itemize}
    \item Approximately 400 stores with 10 cash registers each
    \item For every cash register, a payment every 2 minutes has to be processed
\end{itemize}

Therefore, it is possible to estimate the performance needed for each store to be around 0.083 TPS and the dedicated throughput needed for an application of this type to be:
$$0.083\;TPS\cdot400\approx33\;TPS$$
The tests whose results will be shown below were performed by trying to process 200 transactions for each execution.

\subsection{Lightning Network}
\subsubsection{Development effort and costs}
Each Lightning Network (LN) user must have his own node, in order to manage channels and make payments. At the time of writing, current commercial mobile LN wallets (e.g., \textit{Eclair Mobile}) include a running instance of a LN client. Then, since each LN node instance must also be connected to a Bitcoin node, usually the application is connected to a remote Bitcoin node instance. This is a trade-off needed to balance privacy and usability of the application, by avoiding the need for the user to have enough storage space and computational resources to maintain a local Bitcoin node.
\par A possible architecture for the system could be designed as follows.
\begin{center}
    \includegraphics[width=0.65\textwidth]{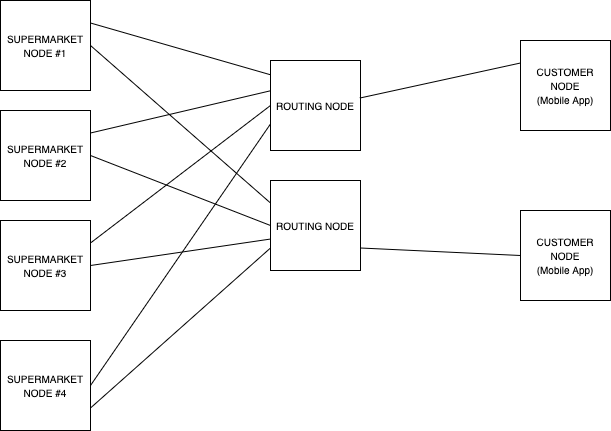}
\end{center}
\par In this context, lines between nodes represent open channels. Routing nodes are third-party \textit{hubs} that are willing to offer routing as a service.
\par This schema, not implying a direct LN channel between grocery shop nodes and each customer, would allow the supermarket to close a limited number of channels (those open with the routing nodes) in order to withdraw funds to Bitcoin L1: the reduced number of channels to be closed mitigates costs for withdrawals and avoids the risk to congest the L1 (when closing channels); as an example, the supermarket could decide to close channels and reopen them periodically (once a week, or when the received amount is above a predefined threshold). 
\par Summarizing, the development of such a system running on top of Lightning Network would need:
\begin{itemize}
    \item A set of Bitcoin nodes and Lightning Network nodes maintained by the supermarket, to receive payments
    \item A mobile wallet application to allow users to make payments
    \item Back-end services capable of generating invoices and verifying if a Lightning Network payment was successful, in order to conclude the transaction with the customer
\end{itemize}
Speaking about transfer fees, in this case they apply since payments by users have to pass through third party routing nodes. As in the case, today, with credit card fees, it is reasonable to assume that these costs would be paid by the supermarket (by discounting the amount requested to the user). To quantify these costs, data from \textit{1ML.com} can be considered: the service estimates the average fee for every routing node that a transfer has to go by to be around 1 satoshi (which is currently equivalent to $\sim$0.00035 USD).
\subsubsection{Customers usability}
\par Current LN wallet applications require a minimum knowledge regarding channels and their opening and closing mechanisms. This procedure could be made semi-automatic with a mobile app dedicated to supermarkets, for example with a user wizard to open a channel with a LN \textit{hub} used by the system; nevertheless, the customer still has to be conscious of what he/she is doing, and the costs related to every operation. Regarding costs, what the customer must pay \textit{on-chain} are the fees to open and close the channel with the routing nodes. At the time of writing, costs for users are the following:
$$Opening\;Channel\;Transaction\;size\approx235\;bytes$$
$$Opening\;Channel\;fee\approx235\;bytes\cdot100\;satoshis/byte\approx23,500\;satoshis=0.000235\;BTC$$
$$Closing\;Channel\;Transaction\;size\approx300\;bytes$$
$$Closing\;Channel\;fee\approx300\;bytes\cdot100\;satoshis/byte\approx30,000\;satoshis=0.0003\;BTC$$

These fees would need to be paid by customers every time they want to \textit{top up} their wallet or when they want to withdraw funds to L1. Currently, it is not possible to add funds to an already existing channel: this means that, if the channel between the user and the routing node runs out of funds, it needs to be closed and reopened again. Some projects are trying to solve this problem (e.g. \textit{Lightning Loop}), but they are still in a beta version.

\subsubsection{Performance}

The implemented architecture is the one described before, using a single routing node between customers node and supermarket node. On Lightning Network Testnet / Bitcoin Testnet (with \textit{lnd} as Lightning node implementation) the obtained results are shown below.\footnote{Details about the performed tests and the proof of concept source code can be found at \url{https://github.com/CosimoSguanci/Blockchain-Layer-2-Proof-of-Concept-App-Polimi}}

\begin{table}[ht]
\begin{tabular}{|c|c|c|c|c|c|c|c|c|}
\cline{1-4} \cline{6-9}
\textbf{Date} & \textbf{Time} & \textbf{TPS} & \textbf{\begin{tabular}[c]{@{}c@{}}Latency \\ (ms)\end{tabular}} &                       & \textbf{Date} & \textbf{Time} & \textbf{TPS} & \textbf{\begin{tabular}[c]{@{}c@{}}Latency \\ (ms)\end{tabular}} \\ \cline{1-4} \cline{6-9} 
20-05-2021    & 08:00 AM      & 60.46        & 943.79                                                           &                       & 05-06-2021    & 10:00 AM      & 66.77        & 903.93                                                           \\ \cline{1-4} \cline{6-9} 
20-05-2021    & 09:00 AM      & 63.96        & 970.2                                                            &                       & 05-06-2021    & 12:00 PM      & 54.36        & 911.66                                                           \\ \cline{1-4} \cline{6-9} 
20-05-2021    & 04:00 PM      & 73.80        & 1095.40                                                          & \multicolumn{1}{l|}{} & 05-06-2021    & 06:00 PM      & 75.38        & 1128.67                                                          \\ \cline{1-4} \cline{6-9} 
20-05-2021    & 05:00 PM      & 51.13        & 888.99                                                           & \multicolumn{1}{l|}{} & 05-06-2021    & 07:00 PM      & 61.18        & 873.11                                                           \\ \cline{1-4} \cline{6-9} 
20-05-2021    & 09:00 PM      & 57.45        & 904.96                                                           &                       & 05-06-2021    & 10:00 PM      & 68.99        & 1075.01                                                          \\ \cline{1-4} \cline{6-9} 
20-05-2021    & 09:30 PM      & 66.49        & 1062.9                                                           &                       & 05-06-2021    & 10:30 PM      & 60.62        & 963.51                                                           \\ \cline{1-4} \cline{6-9} 
\end{tabular}
\end{table}
\newpage
On average, the throughput and transaction latency observed are the following.

\begin{center}
\begin{tabular}{ |c|c| } 
\hline
\textbf{Latency} & \textbf{Throughput} \\
\hline
976.84 ms & 63.38 TPS \\
\hline
\end{tabular}
\end{center}

``Benchmark'' tests were able to process around 63 TPS on average. This result is undoubtedly encouraging, because it shows that the technology has the potentiality to support the presented kind of application, also considering transaction latency: a confirmation waiting time that is less than 2 seconds makes cryptocurrencies a suitable payment method for most real-world scenarios.

\subsection{Plasma}
In order to implement the proof of concept application on Plasma, \textit{Polygon} has been used; it is an adapted account-based implementation of Plasma, enriched with a PoS checkpoint layer.
\subsubsection{Development effort and costs}
Since \textit{Polygon} runs on top of the EVM, in principle each service which runs on an Ethereum-like blockchain is compatible with it: it is enough to change the RPC connection to a node of the Plasma network. In practice, this kind of solution can benefit from the vast Ethereum ecosystem, in terms of libraries (e.g., \textit{web3.js}) and resources.
\par Regarding blockchain nodes, although some services offer RPC APIs, the most reliable solution would be to setup a set of dedicated Plasma chain nodes to better support the application and also avoid traffic limits, which are usually a constraint of third party services. 
\par The necessary architecture for the application would actually need less components than the one described for Lightning Network, as shown in the following picture. \\
\begin{center}
    \includegraphics[width=0.6\textwidth]{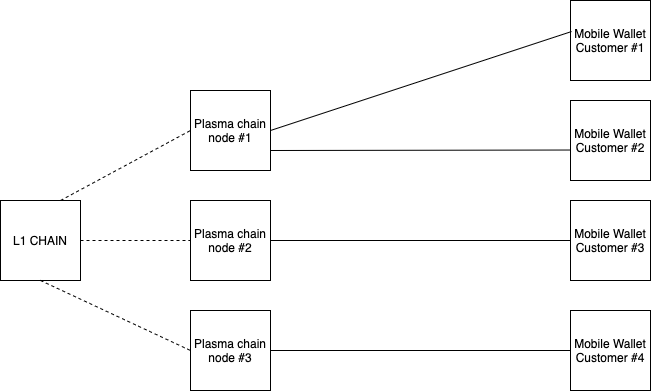}
\end{center}
Here, solid lines represent simple HTTP/WebSocket connections, while dashed lines are merkle commitment of blocks to the L1. The application would consist in an Ethereum-like mobile wallet, to allow users to send transactions on the L2 chain.
\subsubsection{Customers usability}
As stated in previous sections, since the goal is to achieve a decentralized solution, the platform should be \textit{non-custodial}, i.e. only the user has his own private keys. Therefore, the user must be able to securely keep credentials to avoid the risk of losing money. Then, since this application would be specific to the supermarket use case, transfers should be easy to carry out and, therefore, accessible to all customers (this is a problem that should be solved by a proper UX/UI).
\par Regarding costs, payments on a Plasma chain are not free, but are usually very convenient with respect to L1 transactions, due to the different consensus mechanism. At the time of writing, the average gas price on \textit{Polygon} is 3 Gwei: cost for a transfer (that is independent from the transacted amount) is computed as follows:
$$Native\;currency\;transfer\;gas\;needed=21,000$$
$$Average\;Gas\;Price\approx3\;Gwei=3,000,000,000\;Wei$$
$$Transaction\;cost=21,000\cdot3,000,000,000\;Wei=0.000063\;in\;native\;currency$$
These fees are needed for every transaction on the Plasma chain. In the context of the supermarket proof of concept, the supermarket could take charge of these costs by discounting the amount owed by the user. 
\par Fees on L1 are needed only in case of deposit/withdrawal between L1 and L2 and are shown below (considering 40 Gwei as ``standard'' gas price, taken from \textit{ETH Gas Station} at the time of writing).
$$Current\;average\;mainnet\;Gas\;Price\approx40\;Gwei=40,000,000,000\;Wei$$
$$Gas\;needed\;for\;ETH\;deposit\;on\;Plasma\approx77,000$$
$$ETH\;Deposit\;cost\approx77,000\cdot40,000,000,000\;Wei=0.00308\;ETH$$
$$Gas\;needed\;for\;ETH\;withdraw\;from\;Plasma\approx245,000$$
$$ETH\;Withdraw\;cost\approx245,000\cdot40,000,000,000\;Wei=0.0098\;ETH$$
In addition to costs, another constraint for withdrawals is the 7 day dispute period that users must wait before being able to retrieve their funds on the root chain.

\subsubsection{Performance}
Tests have been carried out on the \textit{Polygon} \textit{``Mumbai''} testnet, which has Ethereum's \textit{Görli} testnet as L1 root chain, and the native currency has been involved in transfers. Transactions have been ``bundled'' in a single RPC call by exploiting \textit{web3.js BatchRequest} functionality.

\begin{table}[ht]
\begin{tabular}{|c|c|c|c|l|c|c|c|c|}
\cline{1-4} \cline{6-9}
\textbf{Date} & \textbf{Time} & \textbf{TPS} & \textbf{\begin{tabular}[c]{@{}c@{}}Latency \\ (ms)\end{tabular}} & \multicolumn{1}{c|}{} & \textbf{Date} & \textbf{Time} & \textbf{TPS} & \textbf{\begin{tabular}[c]{@{}c@{}}Latency \\ (ms)\end{tabular}} \\ \cline{1-4} \cline{6-9} 
20-05-2021    & 08:00 AM      & 36.49        & 3688.968                                                         & \multicolumn{1}{c|}{} & 05-06-2021    & 10:00 AM      & 35.41        & 3642.1                                                           \\ \cline{1-4} \cline{6-9} 
20-05-2021    & 09:00 AM      & 35.72        & 3579.16                                                          & \multicolumn{1}{c|}{} & 05-06-2021    & 12:00 PM      & 34.78        & 3625                                                             \\ \cline{1-4} \cline{6-9} 
20-05-2021    & 04:00 PM      & 34.87        & 3625.5                                                           &                       & 05-06-2021    & 06:00 PM      & 39.39        & 3698.01                                                          \\ \cline{1-4} \cline{6-9} 
20-05-2021    & 05:00 PM      & 35.18        & 3694.74                                                          &                       & 05-06-2021    & 07:00 PM      & 39.73        & 3625.9                                                           \\ \cline{1-4} \cline{6-9} 
20-05-2021    & 09:00 PM      & 47.82        & 3657.12                                                          &                       & 05-06-2021    & 10:00 PM      & 43.20        & 3714.44                                                          \\ \cline{1-4} \cline{6-9} 
20-05-2021    & 09:30 PM      & 37.44        & 3648.35                                                          &                       & 05-06-2021    & 10:30 PM      & 26.69        & 3664.43                                                          \\ \cline{1-4} \cline{6-9} 
\end{tabular}
\end{table}

\begin{center}
{\renewcommand\arraystretch{1.4} 
\begin{tabular}{ |c|c| } 
\hline
\textbf{Latency} & \textbf{Throughput} \\
\hline
3,625.31 ms & 37.22 TPS \\
\hline
\end{tabular}}
\end{center}

This case refers to transaction finality on L2. In \textit{Polygon}, finality on-chain (merkle root commitments) is achieved periodically at intervals ranging from 15 minutes to 1 hour. 
\par The obtained performance is worse than Lightning Network. Some possible explanations for this result include the fact that the \textit{Polygon} node client (which is based on \textit{Geth}) may need further optimizations in order to stably support very low block times ($\sim$2 seconds on \textit{Polygon}). Moreover, in this context performance heavily depends on the RPC service provider (for this proof of concept, \textit{Infura} and \textit{BlockVigil} were used). 
\par Anyway, these results still seem to be compatible with the supermarket use case, although better latency is desirable. Considering that the platform used for this test is a Plasma chain which is \textit{shared} with other applications, it is possible to envision a scenario of a dedicated Plasma chain for the supermarket use case, which would lead to better performance (at a cost of increased setup effort and maintenance costs from the supermarket perspective).

\subsection{Rollups}
To showcase the potentiality of Rollups technology, \textit{zkSync}, a ZK Rollup solution proposed by \textit{Matter Labs}, has been used. 
\subsubsection{Development effort and costs}
At the time of writing, \textit{zkSync} is not EVM-compatible (an EVM-based programming model is scheduled to be introduced with \textit{zkSync 2.0}). For this reason, existing software interacting with the Ethereum ecosystem cannot be turned into a \textit{zkSync} application by simply changing a RPC endpoint, and \textit{zkSync} offers libraries supporting the most popular programming languages. The architectural design in this context is easier than the previous two: at the time of writing, becoming a validator of the platform does not seem possible, hence the mobile application would communicate to the \textit{rollups} platform by making use of \textit{zkSync} libraries and APIs. Validators will then publish the validity proof \textit{on-chain}.
\subsubsection{Customers usability}
\label{sec:UseCaseCustomersUsabilityRollups}
From a usability perspective, some common concepts with respect to Lightning Network and Plasma are present, such as those of deposits and withdrawals. To keep the system fully decentralized, users should be able to move their funds to the \textit{rollup} contract \textit{on-chain}, in order to be able to use them on L2.
\par Transaction costs are dramatically lower than L1. Considering Ethereum mainnet as L1 and \textit{zkSync} as L2, $\sim$8x cheaper transfer transactions on L2 (performing a conservative analysis by using the minimum gas price provided by \textit{ETH Gas Station}) can be observed. A comparison of minimum fees (at the time of writing) for transfers on \textit{zkSync} and Ethereum mainnet follows:
$$Current\;mainnet\;ETH\;transfer\;minimum\;fee\approx0.000567\;ETH$$
$$Current\;zkSync\;ETH\;transfer\;minimum\;fee\approx0.0000736\;ETH$$
\par Minimum fee data for \textit{zkSync} was gathered by using the JSON RPC method \verb|get_tx_fee|, offered by \textit{zkSync} APIs.
\par Moreover, through the use of \textit{Batched Transactions}, bundles including many transactions were created, with fees paid by a single sender. This feature fits in particular with the supermarket use case, since an ecosystem in which the supermarket pays for transfer fees (as in the case of traditional payment methods) can be envisioned. In particular, 10 transfer transactions were bundled and executed, with $\sim$0.0001084 ETH paid in fees by a single account.
\par With respect to deposits and withdrawals, the former only requires an \textit{on-chain} fee, while the latter implies an \textit{off-chain} fee, which at the time of writing can be quantified as $\sim$0.0029 ETH. This fee needs to be paid by customers when they want to exit the L2 and return to the root chain. \textit{On-chain} fees for deposits are quantified below (considering ETH as currency and 27 Gwei as the ``standard'' gas price):
$$Gas\;used\;for\;ETH\;deposit\approx62,500$$
$$Current\;ETH\;deposit\;minimum\;fee\approx62,500\cdot27\;Gwei=0.0016875\;ETH$$
\par ZK Rollups allow for $\sim$10 minutes withdrawal times, which, if compared to Plasma and Optimistic Rollups dispute period, is a huge improvement in usability from the user perspective. Another advantage in usability is given by the fact that transfer fees can be paid with the same token that is being transferred, thus eliminating the constraint of having the native currency of the chain to pay fees, as it happens on L1 and Plasma.

\subsubsection{Performance}
In the performed tests, the \textit{Rinkeby} Ethereum testnet was used (transactions here are represented by ETH transfers). Similar to what was done for Plasma (\textit{Polygon}), finality is considered on L2; with high volumes, the proof time generation is expected to be around 10 minutes, i.e. every 10 minutes a validity proof is posted \textit{on-chain} and the transactions included in it can be considered finalized.

\begin{table}[ht]
\begin{tabular}{|c|c|c|c|l|c|c|c|c|}
\cline{1-4} \cline{6-9}
\textbf{Date} & \textbf{Time} & \textbf{TPS} & \textbf{\begin{tabular}[c]{@{}c@{}}Latency \\ (ms)\end{tabular}} & \multicolumn{1}{c|}{} & \textbf{Date} & \textbf{Time} & \textbf{TPS} & \textbf{\begin{tabular}[c]{@{}c@{}}Latency \\ (ms)\end{tabular}} \\ \cline{1-4} \cline{6-9} 
20-05-2021    & 08:00 AM      & 23.19        & 2822.66                                                          & \multicolumn{1}{c|}{} & 05-06-2021    & 10:00 AM      & 3.08         & 2859.34                                                          \\ \cline{1-4} \cline{6-9} 
20-05-2021    & 09:00 AM      & 19.18        & 2816.14                                                          & \multicolumn{1}{c|}{} & 05-06-2021    & 12:00 PM      & 2.26         & 2779.62                                                          \\ \cline{1-4} \cline{6-9} 
20-05-2021    & 04:00 PM      & 34.17        & 2888.18                                                          &                       & 05-06-2021    & 06:00 PM      & 2.86         & 2755.37                                                          \\ \cline{1-4} \cline{6-9} 
20-05-2021    & 05:00 PM      & 25.49        & 2810.06                                                          &                       & 05-06-2021    & 07:00 PM      & 3.19         & 2793.85                                                          \\ \cline{1-4} \cline{6-9} 
20-05-2021    & 09:00 PM      & 24.92        & 2682.78                                                          &                       & 05-06-2021    & 10:00 PM      & 3.78         & 2887.31                                                          \\ \cline{1-4} \cline{6-9} 
20-05-2021    & 09:30 PM      & 18.11        & 3001.98                                                          &                       & 05-06-2021    & 10:30 PM      & 2.98         & 2963.27                                                          \\ \cline{1-4} \cline{6-9} 
\end{tabular}
\end{table}

\begin{center}
{\renewcommand\arraystretch{1.4} 
\begin{tabular}{ |c|c| } 
\hline
\textbf{Latency} & \textbf{Throughput} \\
\hline
2,838.38 ms & 13.6 TPS \\
\hline
\end{tabular}}
\end{center}

Using this kind of solution, a great level of variability in terms of TPS, with respect to the other tested L2 solutions, was observed. The reason could be due to multiple causes: for sure, one of the main differences of this solution compared to LN and \textit{Polygon} (from the point of view of an application designer) is the fact that \textit{zkSync} APIs must be used, as, at the time of writing, it seems to be the only way to interact with the system; this ``constraint'' can lead performance to be highly dependent on some factors, such as APIs rate limits, or simply changes on the back-end side, released by the solution designers. Moreover, some inherent limitations are present, such as the fact that, at the moment of writing, transaction batches allow a maximum of 10 transaction authors per batch.
Therefore, these results should not be considered as the best performance obtainable using a ZK Rollup solution. An interesting fact to notice is that, unlike throughput, transaction latency seems to be stable and also low, especially if compared to the Plasma use case explained in the previous section.
\par To summarize, it is important to point out the fact that improvements for this solution are released very frequently, so anybody can expect it to reach the necessary ``maturity'' to be used reliably, in the context of this proof of concept, in the near future.

\subsection{Use case conclusions}
With the presented proof of concept, a possible practical comparison framework has been shown, both from the point of view of a service provider (the supermarket) and from the perspective of the service users (customers). Finally, regarding performance, it is worth to compare results achieved in previous sections with the current state of one of the most popular blockchain platforms, Ethereum. Running the same benchmark tests on the Ethereum \textit{Ropsten} testnet (which exploits PoW consensus), the following output has been obtained.

\begin{table}[ht]
\begin{tabular}{|c|c|c|c|c|c|c|c|c|}
\cline{1-4} \cline{6-9}
\textbf{Date} & \textbf{Time} & \textbf{TPS} & \textbf{\begin{tabular}[c]{@{}c@{}}Latency \\ (ms)\end{tabular}} &  & \textbf{Date} & \textbf{Time} & \textbf{TPS} & \textbf{\begin{tabular}[c]{@{}c@{}}Latency \\ (ms)\end{tabular}} \\ \cline{1-4} \cline{6-9} 
20-05-2021    & 08:00 AM      & 12.82        & 7984.84                                                          &  & 05-06-2021    & 10:00 AM      & 3.29         & 55804.78                                                         \\ \cline{1-4} \cline{6-9} 
20-05-2021    & 09:00 AM      & 18.8         & 9081.68                                                          &  & 05-06-2021    & 12:00 PM      & 4.37         & 70302.67                                                         \\ \cline{1-4} \cline{6-9} 
20-05-2021    & 04:00 PM      & 6.28         & 33429.86                                                         &  & 05-06-2021    & 06:00 PM      & 5.45         & 41491.75                                                         \\ \cline{1-4} \cline{6-9} 
20-05-2021    & 05:00 PM      & 10.54        & 17026.34                                                         &  & 05-06-2021    & 07:00 PM      & 1.99         & 47401.36                                                         \\ \cline{1-4} \cline{6-9} 
20-05-2021    & 09:00 PM      & 7.135        & 17327.23                                                         &  & 05-06-2021    & 10:00 PM      & 2.12         & 38945.72                                                         \\ \cline{1-4} \cline{6-9} 
20-05-2021    & 09:30 PM      & 2.26         & 23587.22                                                         &  & 05-06-2021    & 10:30 PM      & 5.68         & 38654.89                                                         \\ \cline{1-4} \cline{6-9} 
\end{tabular}
\end{table}

\begin{center}
{\renewcommand\arraystretch{1.4} 
\begin{tabular}{ |c|c| } 
\hline
\textbf{Latency} & \textbf{Throughput} \\
\hline
33,419 ms & 6.73 TPS \\
\hline
\end{tabular}}
\end{center}

In the worst-case scenario for the proof of concept, $\sim$2x improvement in terms of throughput, and $\sim$9x performance improvement for transaction latency, were obtained.
\par In this context, Lightning Network currently seems to be the most mature solution from a performance point of view, since it allows for a substantial throughput and latency improvement. Anyway, it must be said that, as shown, probably other tools require less effort from an architectural/software engineering point of view.
\par The last necessary point to highlight is that all the tools that have been used are under development, therefore their features may change and/or improve over time.

\newpage

\section{Conclusion}
The proposed work has introduced and commented blockchain scalability problem and the most popular solutions to it; in particular, Lightning Network (for Bitcoin) and Raiden Network, Plasma, ZK Rollups and Optimistic Rollups (for Ethereum) have been analyzed, both under their theoretical aspects and five comparison measures (namely scalability, security, decentralization, privacy, fees and micropayments). In order to recap, payment channels solutions, such as LN and RN, are able to achieve highly private transactions, in addition to big improvements to throughput (at least in theory); Plasma, instead, is an extension of traditional sidechains with improvements from the point of view of security and decentralization, and therefore shows interesting properties, especially because its main concepts are not so far from blockchain traditional ones (while state channels proposal, for example, is quite far from them); finally, Rollups introduce the idea of compression (opposed, in a certain sense, to that of reporting \textit{on-chain} just a summary of many \textit{off-chain} transactions), which is an important element for scalability improvements. Moreover, the mentioned Layer 2 technologies could also be used simultaneously on the same platform: Lightning Network on top of Plasma is a possible architecture of this type discussed by Buterin and Poon in the original Plasma proposal.
\par After the theoretical and comparative analysis, the supermarket use case has been described, in order to show a possible scenario of massive adoption of Layer 2 solutions (especially related to micropayments): the obtained results show how, under latency and throughput points of view, all the frameworks perform better than a ``pure'' L1 application (considering Bitcoin and Ethereum). \par Lightning Network constitutes the most mature technology for scaling Bitcoin and, in spite of the \textit{hub and spoke} current architecture that can partly menace some of its theoretical features, it will surely play a crucial role for blockchain massive adoption. \par For what regards Ethereum, instead, the reasoning is a bit more complicated, because the community, in addition to Layer 2 proposals, is heavily pushing on Eth2 development as main scaling solution. Ethereum 2.0 (Eth2) is ``a set of upgrades that improve the scalability, security, and sustainability of Ethereum''\footnote{Quotation from \url{https://ethereum.org/en/eth2/}} by introducing proof-of-stake (PoS) and sharding; in terms of throughput, developers' goal is that of achieving thousands of transactions per second.\footnote{Going into Ethereum 2.0 details is not one of the goals of this work; the interested reader can find useful information at \url{https://ethereum.org/en/eth2/} and \url{https://docs.ethhub.io/ethereum-roadmap/ethereum-2.0/eth-2.0-phases/}} The concern, now, is to understand whether Eth2 will cause uselessness of all Ethereum L2 solutions presented in this work; in order to face the problem, it is important to point out that Ethereum 2.0 is estimated to be ready for full launch in more than one year (at the moment of writing) and that, in any case, Layer 2 technologies can integrate on top of it, as well; so, adapted implementations of Plasma and Rollups are likely to be the most important scaling solutions for some time, then they will continue to have a key role for higher throughput (in general, but especially in the early phases of Eth2 adoption). As a matter of fact, as stated by Vitalik Buterin ``the scalability gains from the Layer 1 improvements and Layer 2 improvements do ultimately multiply with each other''. Therefore, it is possible to consider sharding and existing L2 solutions as complementary, as performance improvements add up to each other.
\par Raiden Network, instead, is still an immature technology (at the moment of writing) and, according to most recent estimations, will probably have few time to settle up before Ethereum updates; in any case, it would offer unique privacy features that users might look for when transacting: this fact is probably going to lead to RN adoption inside Ethereum ecosystem, independently of throughput and Eth2 achievements. \par In conclusion, Layer 2 solutions certainly represent the present and future for solving blockchain scalability problem, without renouncing to a secure and decentralized architecture.

\newpage
\nocite{*}
\printbibliography
\end{document}